\documentclass[12pt, preprint]{aastex}

\usepackage{wasysym} 
\usepackage{amssymb}
\usepackage{lscape}
\usepackage{color}

\begin{document}

\title{The Evolution of Quasar \ion{C}{4} and \ion{Si}{4} Broad Absorption Lines Over Multi-Year Time Scales}

\author{Robert R. Gibson\altaffilmark{1,2}, W.~N. Brandt\altaffilmark{2}, S.~C. Gallagher\altaffilmark{3}, Paul C. Hewett\altaffilmark{4}, Donald P. Schneider\altaffilmark{2}}
\email{rgibson@astro.washington.edu}

\altaffiltext{1}{Department of Astronomy, University of Washington, Box 351580, Seattle, WA 98195, USA}
\altaffiltext{2}{Department of Astronomy and Astrophysics, Pennsylvania State University, 525 Davey Laboratory, University Park, PA 16802, USA}
\altaffiltext{3}{Department of Physics and Astronomy, The University of Western Ontario, 1151 Richmond Street, London, ON N6A 3K7, Canada}
\altaffiltext{4}{Institute of Astronomy, University of Cambridge, Cambridge CB3 0HA, UK}

\shorttitle{BAL Variation on Multi-Year Time Scales}
\shortauthors{Gibson et al.}

\hbadness=10000

\clearpage

\begin{abstract}
We investigate the variability of \ion{C}{4}~$\lambda$1549 broad absorption line (BAL) troughs over rest-frame time scales of up to $\approx$7~yr in 14~quasars at redshifts $z \ga 2.1$.  For 9 sources at sufficiently high redshift, we also compare \ion{C}{4} and \ion{Si}{4}~$\lambda$1400 absorption variation.  We compare shorter- and longer-term variability using spectra from up to four different epochs per source and find complex patterns of variation in the sample overall.  The scatter in the change of absorption equivalent width (EW), $\Delta$EW, increases with the time between observations.  BALs do not, in general, strengthen or weaken monotonically, and variation observed over shorter ($\lesssim$months) time scales is not predictive of multi-year variation.  We find no evidence for asymmetry in the distribution of $\Delta$EW that would indicate that BALs form and decay on different time scales, and we constrain the typical BAL lifetime to be $\gtrsim$30~yr.  The BAL absorption for one source, LBQS~0022+0150, has weakened and may now be classified as a mini-BAL.  Another source, 1235+1453, shows evidence of variable, blue continuum emission that is relatively unabsorbed by the BAL outflow.  \ion{C}{4} and \ion{Si}{4} BAL shape changes are related in at least some sources.  Given their high velocities, BAL outflows apparently traverse large spatial regions and may interact with parsec-scale structures such as an obscuring torus.  Assuming BAL outflows are launched from a rotating accretion disk, notable azimuthal symmetry is required in the outflow to explain the relatively small changes observed in velocity structure over times up to 7~yr.
\end{abstract}


\section{INTRODUCTION\label{introSec}}

High-velocity, structured outflows in the central regions of quasars (QSOs) generate broad absorption lines (BALs) from material along the line of sight to the QSO ultraviolet (UV) emitter.  These ``BAL outflows'' are launched, ionized, and accelerated by physical processes in the heart of the QSO, and can provide an important means of carrying material and energy out of the QSO's central region.  BAL features can reach high outflow velocities (30,000~km~s$^{-1}$ or more), and the absorption signatures of these outflows are, by definition, at least 2000~km~s$^{-1}$ wide \citep{wmfh91}, although BALs can be much broader.  BALs are observed in up to $\approx$15\% of QSOs in optical/UV spectroscopic surveys \citep[e.g.,][and references therein]{wmfh91, hf03, rrhsvfykb03, thrrsvkafbkn06, gjbhswasvgfy09}, and the fraction of QSOs with BALs rises to 17--23\% after correcting for optical/UV selection effects \citep[e.g.,][]{hf03, ksgc08, gjbhswasvgfy09}.  The intrinsic fraction may be $\sim$2 times higher if less-restrictive criteria are used to select BALs \citep[e.g.,][]{thrrsvkafbkn06, dss08, gb08}; in particular, narrower, mini-BAL absorption features may also be related to BALs \citep[e.g.,][]{akdjb99, gbcg02, gbgs09}.  Some physical models associate BAL outflows with equatorial accretion disk winds that may be ubiquitous in QSOs \citep[e.g.,][]{mcgv95, psk00}.  BAL outflows may also result during evolutionary stages when QSOs expel thick shrouds of gas and dust \citep[e.g.,][]{vwk93, gbd06}.

High-resolution spectroscopic studies have found that BAL absorption is frequently saturated, but does not completely obscure the continuum emitter.  \citet{ablgwbd99} concluded that BAL absorption troughs are generally shaped by the outflow kinematic structure and geometry, with the depth of the absorption feature representing the fraction of the continuum emitter covered by one or more components of the absorbing outflow at a given velocity.  In this model, studies of BAL variability primarily reveal ongoing changes in the structure of the outflow.  Variability studies have also sought to test the ``covering factor'' hypothesis by identifying correlations between absorption and emission variation that would indicate that absorption is not strongly saturated, but that the absorption strength is related to the photoionization-dependent column density of the absorbing ion.  The results of these studies are mixed, but conclusive evidence for photoionization-dominated variation has not been found in the general population of BALs \citep[e.g.,][]{bjb89, bjbwmk92, lwbhsyvb07, gbsg08G08}.

With sufficient spectroscopic observing resources, BAL monitoring studies could greatly improve our knowledge of the composition and evolving structure of QSO outflows, and the physical mechanisms that launch and accelerate outflows.  In practice, however, BAL variability studies have been limited by several constraints.  Because BAL absorption is most commonly studied in the \ion{C}{4} $\lambda$1549 line, BAL observations must be obtained either in the UV or of sources at sufficiently high redshift to place \ion{C}{4} BALs into a bandpass observable from ground-based telescopes.  Because most known BAL QSOs are at $z \ga 1.5$, it can be difficult to study variation over extended (rest-frame) time scales.  With modern AGN surveys such as the Sloan Digital Sky Survey \citep[SDSS;][]{y+00}, it is now possible to identify thousands of QSOs with observed BAL absorption \citep[e.g.,][]{thrrsvkafbkn06, gjbhswasvgfy09}.  Yet it is necessary to have spectra obtained $\gtrsim 10$~yr apart in order to span several rest-frame years.

In this work, we examine the variation of BAL QSOs observed over rest-frame time scales of up to 7~yr.  In order to construct a reasonably representative sample, we have selected all BAL QSOs at redshifts $z \gtrsim 2.1$ from the catalog of \citet{hf03} that are visible to the Hobby-Eberly Telescope \citep[HET;][]{rabbcfggghkklmrrssss98}.  Our sample therefore inherits the selection properties of the Large Bright Quasar Survey \citep[LBQS;][]{hfc95}.  We re-observed each of the 14~QSOs matching our selection criteria using the HET Low Resolution Spectrograph \citep[LRS;][]{hnmtcm98} in order to obtain spectra with adequate signal-to-noise ratio (S/N) and spectral resolution for comparison to the LBQS epoch.  Additionally, many of the sources in our study have spectra available from intermediate epochs, obtained either as part of the follow-up study by \citet{wmfh91}, or during the course of the SDSS.  We will refer to the former set of spectra as the ``Palomar'' BAL spectra.  The LBQS spectra in our current study and that of \citet[][hereafter G08]{gbsg08G08} were obtained in the years 1986--1989.  With the addition of the Palomar and SDSS spectra, we can sample time scales of months to several years for individual sources.

In general, BALs vary in complex ways over multi-year time scales (G08).  While studies of individual BAL QSOs are needed to understand how BALs appear \citep[e.g.,][]{hkrph08, lhcg09}, disappear \citep[e.g.,][]{jsbbchl01}, and accelerate \citep[e.g.,][]{hsher07}, it is also important to determine the general characteristics of BAL outflow evolution.  In the current work, we take another step in that direction by comparing \ion{C}{4} BAL absorption profiles in up to four epochs spanning a range of time scales from months to 7~yr.  Many of our sources are at sufficiently high redshift to shift the \ion{Si}{4} absorption region into the optical bandpass, enabling a study of the simultaneous variation of absorption from two different ions (\ion{Si}{4} and \ion{C}{4}).

\subsection{Conventions\label{conventionsSec}}

Because conventions vary among studies, we briefly describe the terminology used in this work.  We define outflow velocities to be negative (with respect to QSO emission rest frames).  Positive velocities would indicate features that are at {\it longer} wavelengths than the wavelength corresponding to (rest-frame) zero velocity.  However, we use the terms ``greater'' and ``smaller'' velocities to refer to the {\it magnitude} of the velocity, so that an outflow speed of \mbox{--10,000~km~s$^{-1}$} is ``greater'' than a speed of \mbox{--5000~km~s$^{-1}$}.  We consider absorption equivalent widths (EWs) to be negative, so that a negative change in $\Delta$EW between two epochs corresponds to a strengthening BAL.  Errors in EW are calculated by propagating standard errors, with an estimate of 10\% error in the continuum.

Although we often quote the averaged doublet wavelengths to identify a spectral line, we perform absorption outflow calculations using the wavelength of the red component of the doublet.  In effect, we use absorption rest wavelengths of 1550.77~\AA\ for \ion{C}{4} and 1402.77~\AA\ for \ion{Si}{4}.  This convention was adopted because the onset of a strong absorption trough from outflowing material would be associated with the red side of the doublet.  Unless otherwise noted, wavelengths in this work refer to rest-frame values.  We use vacuum wavelengths for all epochs.

Sources identifiers that are formatted like ``0009+0219'' refer to LBQS source designations using B1950.0 coordinates.  Throughout, we use a cosmology in which $H_0 = 70$~km~s$^{-1}$~Mpc$^{-1}$, $\Omega_M = 0.3,$ and $\Omega_{\Lambda} = 0.7$.

\section{OBSERVATIONS AND DATA REDUCTION}
\subsection{Observations and Methods\label{dataRedSec}}

Our sample is comprised of the 14~BAL QSOs at $z \gtrsim 2.1$ identified in the LBQS survey by \citet{hf03} with \ion{C}{4} absorption regions that can be observed by the HET.  We selected sources from the \citet{hf03} list with redshifts $z \gtrsim 2.1$ so that potential high-velocity absorption extending blueward to 1400~\AA\ could be observed in the HET LRS bandpass (observed-frame wavelengths $\ga$4340~\AA).  One source, LBQS~2212--1759, was excluded because its declination was lower than the HET limit.  The observations used in our study are listed in Table~\ref{obsLogTab}.

LBQS spectra were obtained from 1986--1988 with the Multiple Mirror Telescope (MMT) spectrograph.  They cover a wavelength range of 3200--7400~\AA\ and generally have a spectral resolution of $\sim$6~\AA\ with wavelength calibrations accurate to $\sim$4~\AA\ \citep{fchmtwa87}.  The ``Palomar'' spectra were taken as part of a campaign to obtain spectra of LBQS QSOs, especially BAL QSOs \citep{wmfh91}.  Most of the Palomar spectra were obtained using the Hale Telescope with the Double Spectrograph \citep{og82} from 1988--1989 with a resolution of about 6.5~\AA\ in bluer wavelengths.  These spectra are available for most of our sources.  We discarded three cases (0021--0213, 0025--0151, and 0029+0017) for which the original spectrum was not available at rest-frame wavelengths $\lambda < 1550$~\AA.  We also omitted the spectrum for 1231+1320 because of apparent inconsistencies in its flux correction at bluer wavelengths.  SDSS spectra for four sources were obtained from the public SDSS Data Release~5 (DR5) database \citep{a-m+07}.  These spectra have a higher resolution ($\sim$3~\AA) than the other spectra in our study.

Since 2007, we have also been observing these QSOs with the HET LRS.  We used the g2 grism and a GG385 filter with an 1.5$\arcsec$ slit, giving wavelength coverage down to $\approx$4340~\AA\ on the blue side and a spectral resolution (estimated from fits to sky lines) of $\approx$6.2~\AA, similar to those of the LBQS and Palomar spectra.  We reduced the spectra using standard IRAF procedures and flux-corrected each spectrum using a standard star observation obtained the same night.  One of our sources, 1235+1453, had already been observed by our group in 2002, so we did not re-observe that source.  Exposure times for our sources ranged from $\approx$15--20 minutes.  In a small number of cases, we interpolated over bins to remove ``glitches'' in the spectra that were much narrower than the instrumental resolution and are typically associated with cosmic ray hits or detector defects.

Throughout this work, we focus on absorption (and emission) profiles obtained by dividing observed spectral features by an estimated model of continuum and broad-line emission.  Given the possible errors in flux calibration for the HET and other observing epochs that are beyond our control, we do not depend on absolute or relative flux calibration.  Additionally, visual inspection indicated that the wavelength calibration between telescopes may have differed slightly for up to 5 sources.  We examined non-BAL absorption and emission in spectral regions surrounding the broad \ion{C}{4} emission line at $1549$~\AA\ to visually determine a multiplicative correction factor ($\la$0.3\% in 4 of 5 cases) that we applied to LBQS wavelengths to obtain a satisfactory match among epochs.  We selected a fiducial common redshift for every epoch of each source from the SIMBAD database;\footnote{\tt http://simbad.u-strasbg.fr/simbad/sim-fid} these values are listed in Table~\ref{obsLogTab}.  These small calibration adjustments are sufficient for our study, which relies upon a relatively small wavelength range (1300--1550 \AA) and does not require precise absolute wavelength calibration.

For each spectrum in our study, we fit an emission model following the prescription in G08 and \citet{gjbhswasvgfy09}.  In brief, we fit an intrinsically reddened power law continuum model to each spectrum after correcting for Galactic reddening.  We initially fit only wavelength regions that are expected to be relatively free of emission or absorption features.  We then iteratively fit emission lines to \ion{Si}{4}~$\lambda$1400, \ion{C}{4}~$\lambda$1549, \ion{Al}{3}~$\lambda$1857, \ion{C}{3}]~$\lambda$1909, and \ion{Mg}{2}~$\lambda$2799.  At each stage, we exclude regions that deviate strongly from the fit in order to exclude strong absorption and emission features.  We model the cores and wings of each emission line using a single Voigt profile for convenience, although we attach no physical significance to this model.  Our approach to fitting emission line regions is to fit the emission that is present, excepting regions that exhibit obvious absorption features.  If emission lines are absorbed by variable absorbing components that are broad and smooth enough to cover the broad emission line without leaving obvious absorption features, we would not be sensitive to this effect.  Of course, emission lines may also vary for a number of reasons unrelated to absorption.  Also, any variable asymmetry in broad emission line profiles will not be perfectly modeled with our symmetric line profiles.  Our chosen approach is motivated partly by the capabilities of the current data set; future data sets with improved flux calibration and spectral resolution will be able to address these issues in more detail.

\begin{figure} [ht]
  \begin{center}
      \includegraphics[width=5in, angle=-90]{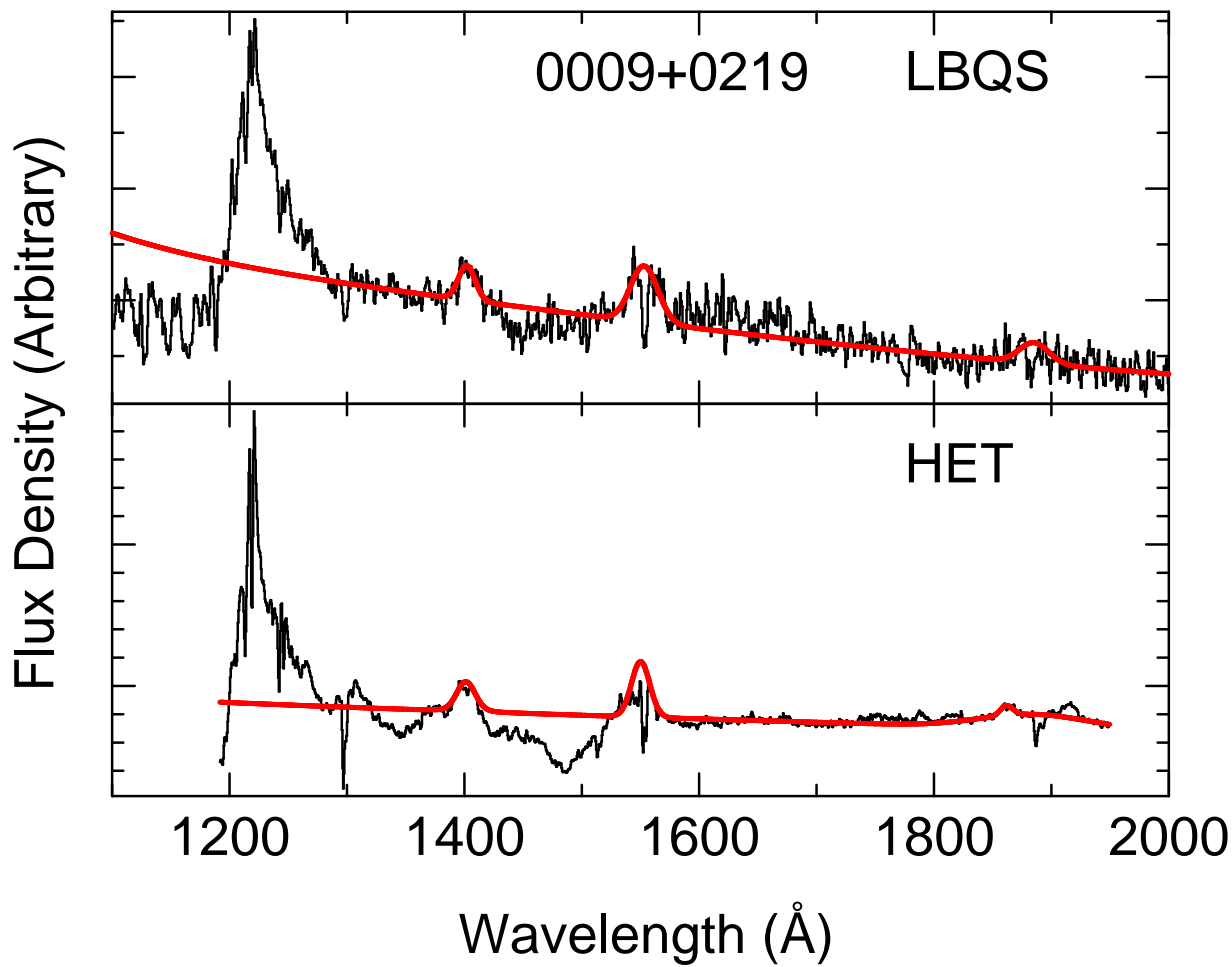}
      \caption{\label{specAndModelFig}LBQS, Palomar, SDSS, and HET spectra (when available) plotted as flux density (erg~s$^{-1}$~cm$^{-2}$~\AA$^{-1}$), with an arbitrary $y$-axis normalization for each panel.  The source name is shown at top (e.g., ``0009+0219'') and observing epochs are listed for each panel.  Spectral properties for each observing epoch and model fitting techniques are described in \S\ref{dataRedSec}.  The spectra have been smoothed with a boxcar filter 3~bins wide for visual clarity.  The thick red line indicates our emission model fit in each case.  Error bars have been omitted for clarity, but are representative of the scatter in each spectrum.  A color version and an extended version of this figure are available in the online journal.}
   \end{center}
\end{figure}

We inspected and manually adjusted the fits in each case, using spectra from other epochs as a guide in a small number of cases where there was ambiguity in continuum placement or emission line shape.  When adjusting continua, we ignored the region from $\sim$2200--3000~\AA\ that can be contaminated by ionized Fe emission; such emission is difficult to constrain and the continuum at those wavelengths does not strongly influence our estimates of \ion{Si}{4} and \ion{C}{4} properties in any case.  The spectra and emission models for each source and epoch are shown in Figure~\ref{specAndModelFig}.

Most of the data analysis for this project was performed using the ISIS platform \citep{hd00}.

\subsection{The Case of 1235+1453\label{1235+1453Sec}}

\begin{figure} [ht]
  \begin{center}
      \includegraphics[width=5in, angle=-90]{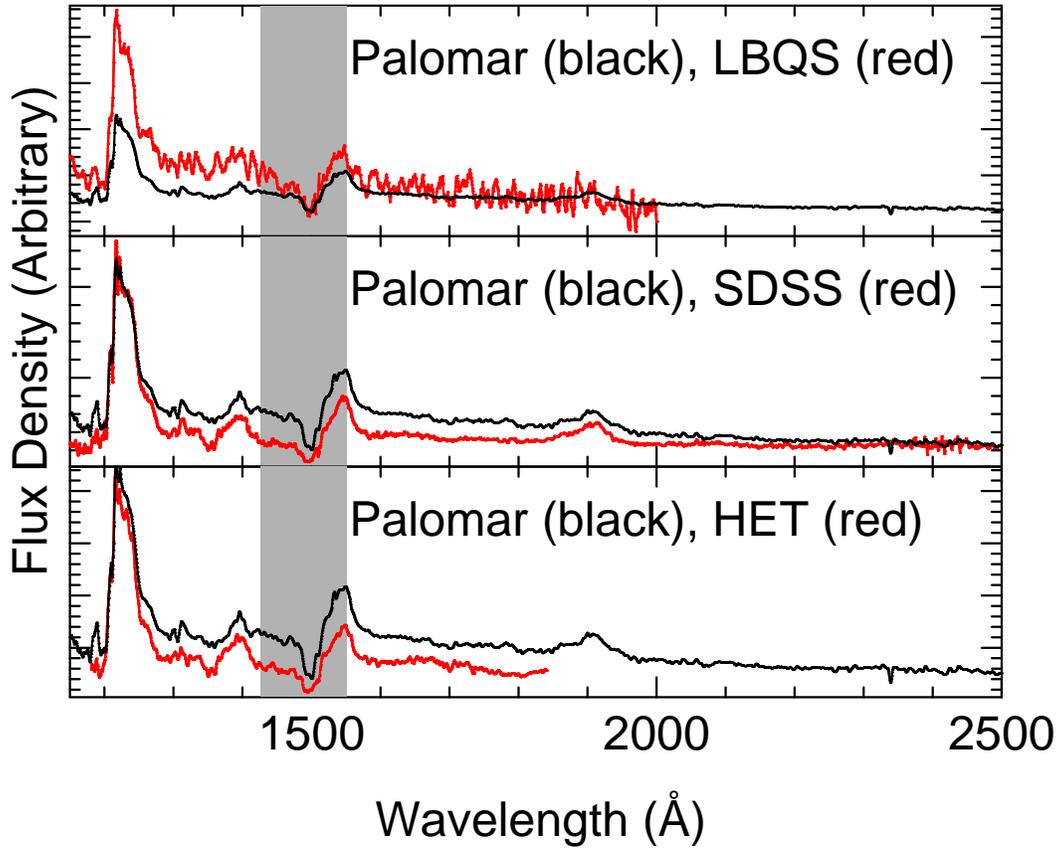}
      \caption{\label{plotCompare1235+1453Fig}Comparison of the spectra obtained for 1235+1453.  The spectra have been smoothed using a boxcar of width~7 bins for visual clarity.  In each panel, the Palomar epoch is shown in black, while the red line represents the LBQS (top), SDSS (middle) and HET (bottom) spectra.  The shaded portion represents the potential \ion{C}{4} BAL absorption region.  A color version of this figure is available in the online journal.}
   \end{center}
\end{figure}

One of the sources, 1235+1453, is an outlier in several of our analyses.  The C IV BAL of this source is dramatically weaker in the Palomar epoch.  Figure~\ref{plotCompare1235+1453Fig} compares the LBQS, SDSS, and HET spectra to the Palomar spectrum of 1235+1453.  The figure shows that the continuum of 1235+1453 has also changed dramatically over time; it is much bluer in the earlier (LSST and Palomar) epochs.  This is most clear in the middle panel of Figure~\ref{plotCompare1235+1453Fig}, where we compare the two epochs with the best flux calibration (Palomar and SDSS).

\begin{figure} [ht]
  \begin{center}
      \includegraphics[width=5in, angle=-90]{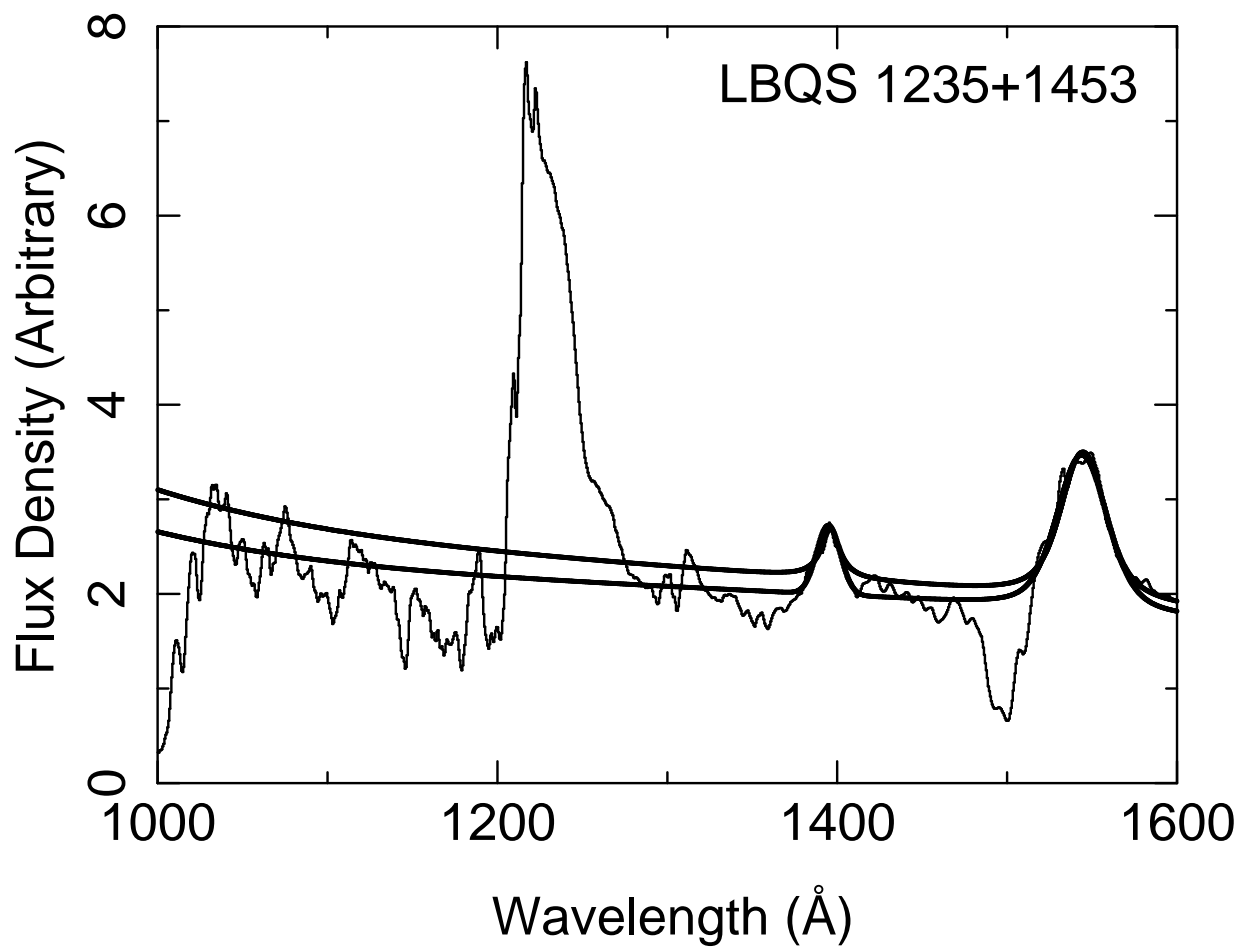}
      \caption{\label{compare1235+1453ModelsFig}The Palomar spectrum of 1235+1453 (binned as in Figure~\ref{plotCompare1235+1453Fig}), with two alternative models for continuum, \ion{Si}{4}, and \ion{C}{4} emission plotted as thick lines.  The top model is the one we use in this work, selected to match BAL absorption in other epochs more closely.  The bottom model attempts to match the continuum underlying the emission line at 1300~\AA, but falls below the emission in the 1400--1450~\AA\ range and in the absorbed region blueward of 1200~\AA.  See \S\ref{1235+1453Sec} for further discussion.}
   \end{center}
\end{figure}

The values of the \ion{C}{4} BAL EW at the different epochs depend on the continuum level in each spectrum.  We have experimented with different continuum fits to model the Palomar-epoch BAL absorption and are unable to find a satisfactory model that matches the apparent blue continuum and also results in a \ion{C}{4} BAL resembling those in other epochs.  Figure~\ref{compare1235+1453ModelsFig} demonstrates two alternative continuum placements.  For this work, we have opted to use the higher continuum placement.  The lower continuum attempts to match the continuum level ``underneath'' the emission line at 1300~\AA, but also falls below the emission levels at 1400--1450~\AA\ and in the absorbed region blueward of 1200~\AA.  An even higher continuum placement would be required to obtain a BAL strength equivalent to other epochs.

The existing photometry and spectra for 1235+1453 are consistent with the presence of a very blue ultraviolet continuum component that declines over time and is not absorbed by the BAL outflow.  The differences in the Palomar-epoch BAL are due to changes in the absorption trough depth (by 20--25\% of the continuum level), rather than changes in the absorption profile.  A variable, blue continuum component would also explain the substantial changes in the (rest-frame) UV spectral shape of 1235+1453 as well as the fainter $g$-band magnitude at the epoch of the SDSS spectrum relative to that of the Palomar spectrum ($\simeq$0.6\,mag fainter over a rest-frame interval of $\simeq$4~yr).  From the current data, the cause of the observed blue variability is not obvious.  SDSS spectroscopic studies have demonstrated that quasars are generally bluer when brighter \citep{wvkspbrb05}, and the spectral shape variability could be associated with changes in accretion rate \citep{pvthwksb06}.

The amplitude of the variability change shown is unusual but not extreme for a representative quasar over several years \citep[e.g.,][]{vwkabhirsyblns04, wvkspbrb05}.  We also note that 1235+1453 was also somewhat unusual in the study of \citet{gbcpgs06}, in that it showed the maximum observed degree of X-ray weakness out of a sample of 35 LBQS BAL QSOs.  While the variable continuum of quasar spectra has been studied in detail \citep[e.g.,][]{wvkspbrb05}, we believe our discovery that the variable continuum component of 1235+1453 does not show the presence of high-ionization BAL troughs is a new result.  Confirmation of such behavior among the BAL QSO population would provide significant new constraints on the origin of BAL outflows and their geometry with respect to quasar emission regions.

Although the behavior of 1235+1453 is unusual in the context of the BAL QSO sample investigated here, we do not wish to bias our data by removing 1235+1453 from our sample.  We retain 1235+1453 in our study, but we note cases where it is an outlier that significantly affects our results.

\section{ANALYSIS AND DISCUSSION\label{discSec}}

\begin{figure} [ht]
  \begin{center}
      \includegraphics[width=5in, angle=-90]{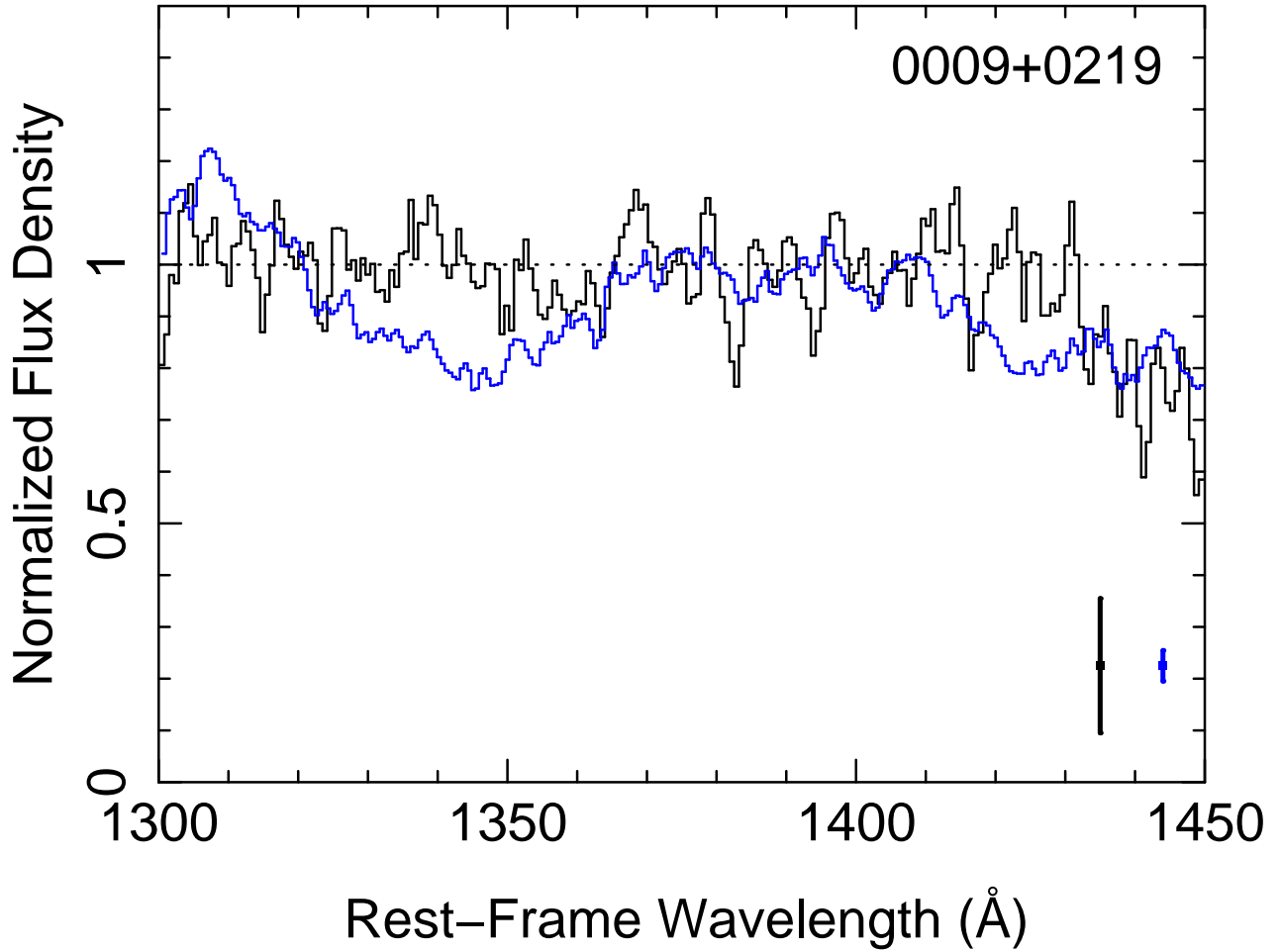}
      \caption{\label{compareSiIVAbs0009Fig}The \ion{Si}{4} BAL absorption regions of LBQS (black), Palomar (red, when available), SDSS (green, when available), and HET (blue) spectra for our sources, with continuum fits divided out.  The spectra have been smoothed by a boxcar filter 3~bins wide for visual clarity.  Typical error bars for the unsmoothed spectra are shown at the bottom right, color-coded for each epoch.  A color version and an extended version of this figure are available in the online journal.}
   \end{center}
\end{figure}

\begin{figure} [ht]
  \begin{center}
      \includegraphics[width=5in, angle=-90]{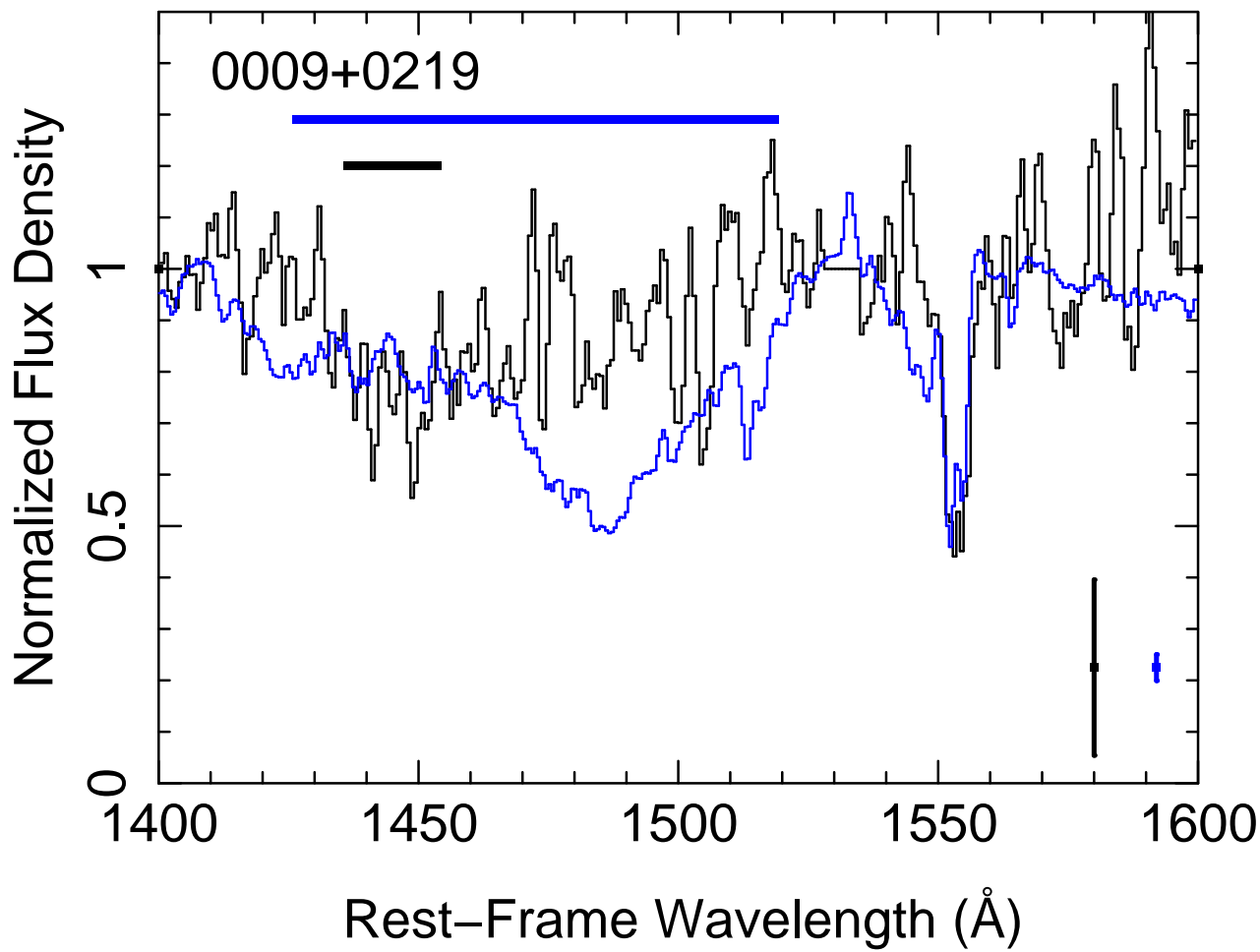}
      \caption{\label{compareCIVAbs0009Fig}Same as Figure~\ref{compareSiIVAbs0009Fig}, but for the \ion{C}{4} absorption region.  Thick horizontal lines (color-coded to match the spectra) indicate regions that are formally identified as components of a \ion{C}{4} BAL in the calculation of BI$_0$ (defined in \S\ref{fracVarEWSec}).  Because the formal definition of the balnicity index only extends out to outflow velocities $-25,000$~km~s$^{-1}$, no absorption blueward of $\approx$1426~\AA\ is included.  A color version and an extended version of this figure are available in the online journal.}
   \end{center}
\end{figure}

Figure~\ref{compareSiIVAbs0009Fig} and Figure~\ref{compareCIVAbs0009Fig} show the ``ratio spectra'' of the \ion{Si}{4} and \ion{C}{4} absorption regions, respectively, with our emission model divided out.  The figures demonstrate that patterns of BAL variation are, in general, complex.  Rather than attempting to describe the wide variety of changes in these absorption features, we focus instead on statistical statements about the evolution of BAL absorption.  We rely primarily on measures such as the EW that are integrated over the whole wavelength range where absorption could occur.  We prefer such measures to those that are calculated on a set of bins specific to each BAL, such as the ``balnicity index'' \citep[BI;][]{wmfh91} and maximum outflow velocity ($v_{max}$), because these measures are highly dependent on the wavelengths chosen to be the boundaries of each BAL trough.  BAL boundaries can change dramatically due to spectral resolution and random fluctuations in even a single spectral bin, and we do not wish the results of our study to be unduly influenced by our choice of BAL boundaries in each observing epoch.

Of course, the integrated measure of EW is not sensitive to changes in which one part of a BAL weakens, while another part strengthens.  In some parts of our study (e.g., \S\ref{siIVCIVShapeVarSec} and \S\ref{varPatternsSec}), we investigate the patterns of variation in more detail.  Additional statistics could also be employed to measure, e.g., the sum of the magnitude of change in absorption profiles.  Such statistics may be particularly useful in larger-scale studies using spectra having uniformly high resolution and S/N.

Throughout this work, we search for correlations between absorption and emission properties using the Spearman Rank Correlation test.  In addition to describing each test in the work below, we summarize the results of our correlation tests in Table~\ref{corrTab}.  We do not generally draw conclusions from correlations at a confidence level $< 99$\%.

\subsection{Variation in \ion{C}{4} Equivalent Width\label{varEWSec}}

We calculate the rest-frame EW of each absorption region using the continuum and emission-line fits described in \S\ref{dataRedSec}.  To calculate \ion{C}{4} absorption EWs, we integrate over the entire region blueward of the \ion{C}{4} rest wavelength (1550.77~\AA) out to a velocity of --30,000~km~s$^{-1}$.  We integrate EWs in units of Angstroms or km~s$^{-1}$; a conversion of $\approx$200~km~s$^{-1}$ per Angstrom is a good approximation for the \ion{C}{4} absorption region.  Our calculated values are listed in Table~\ref{cIVBALEWTab}.

\begin{figure} [ht]
  \begin{center}
      \includegraphics[width=5in, angle=270]{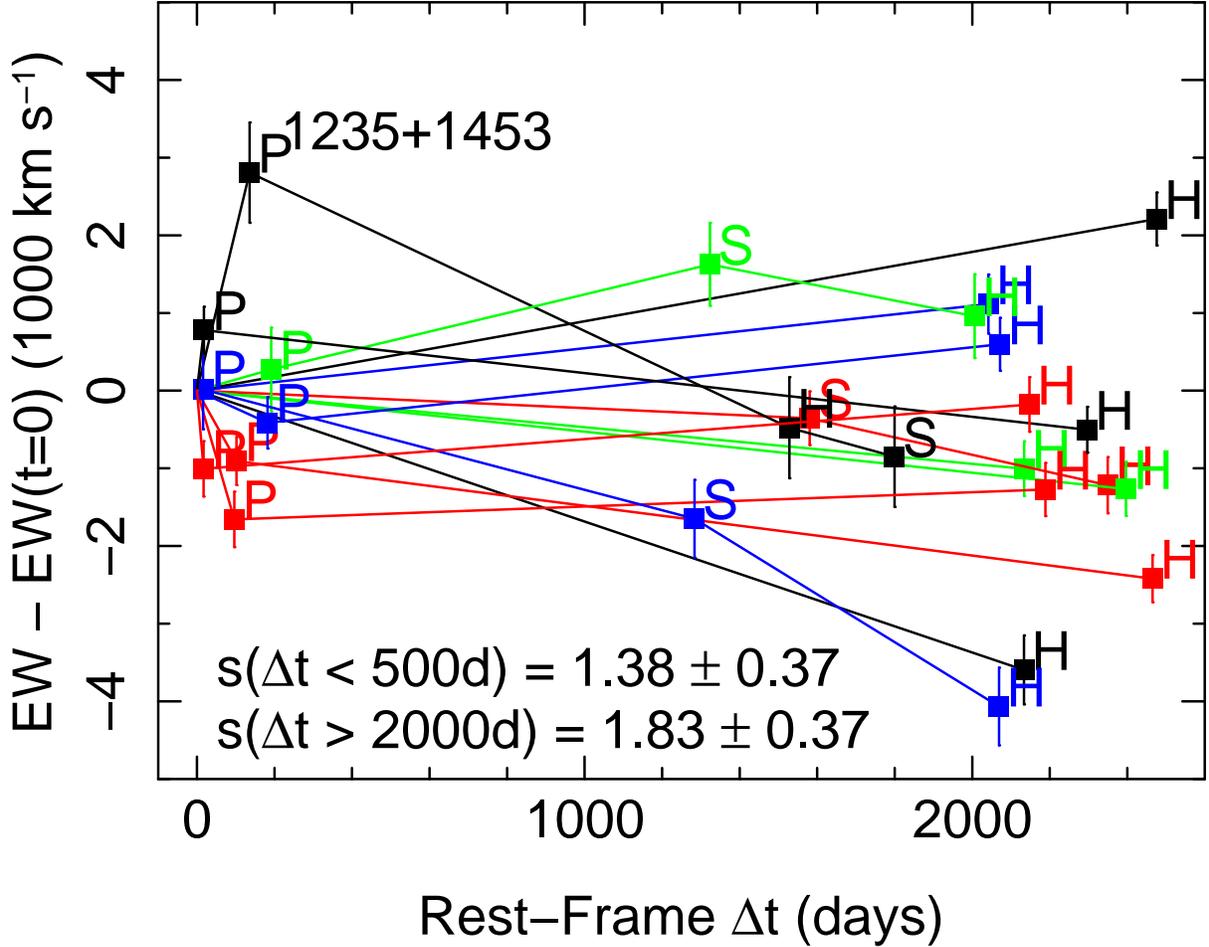}
      \caption{\label{dEWFig}The change in EW over time for sources in our study.  Because absorption EW is negative, strengthening BALs are in the lower half of the plot.  Here, EWs are calculated by integrating over all bins in the velocity range --30,000 to 0~km~s$^{-1}$ from the \ion{C}{4} rest wavelength.  The observing date of the first (LBQS) epoch is defined to be $\Delta t_{sys} = 0$ for each source, and the EW of the first (LBQS) epoch is subtracted from all EWs so that only the change in EW is shown.  Lines connect measurements corresponding to a single source, and each individual measurement is marked with a letter indicating the corresponding observing campaign (``P'' for Palomar, ``S'' for SDSS, and ``H'' for HET).  The square root, $s$, of the unbiased sample variance (in units of 1000~km~s$^{-1}$) is shown for comparison between sources with $\Delta t_{sys} < 500$ and $\Delta t_{sys} > 2000$~days.  In the color version (online), the points and lines for each source have been given different (arbitrary) colors for visual clarity.  A color version of this figure is available in the online journal.}
   \end{center}
\end{figure}

Figure~\ref{dEWFig} shows the evolution of EWs over time.  For each source, we subtracted the observing date and measured EW for the first (LBQS) epoch from those of the following epochs so that the LBQS BALs correspond to the point (0, 0) for each source.  All measurements for a single source are connected with a solid line in the figure, and individual measurements are labeled according to the observing campaign (``P'' for Palomar, ``S'' for SDSS, and ``H'' for HET).

In order to investigate whether the spread of $\Delta$EW increases with time, we calculated $s$, defined as the square root of the unbiased sample variance of $\Delta$EW, for the 8~sources with $\Delta t_{sys} < 500$~days and also for the 13~sources with $\Delta t_{sys} > 2000$~days.  Here, $\Delta t_{sys}$ represents the elapsed time in the rest frame.  For the first set, comprised solely of Palomar observations, $s(\Delta t_{sys} < 500) \approx 1380 \pm 370$~km~s$^{-1}$ and the median time between observations is $96$~d.  For the latter set, $s(\Delta t_{sys} > 2000) \approx 1830 \pm 370$~km~s$^{-1}$ and the median time between observations is $2150$~d.  Here we estimate the variance on $s^2$ as $var(s^2) \approx 2s^4/(N-1)$ for $N$ data points \citep[e.g.,][]{etpvmmdw02}; from this we estimate the error on $s$ as $s /\sqrt{2 (N - 1)}$.

Our measured trend of increasing spread in $\Delta$EW over time is qualitatively consistent with that of G08, although the difference in variance between the LBQS--Palomar and LBQS--HET observations is not highly significant and the variance for the Palomar set is somewhat larger than that measured over similar time scales in G08.  The variance of the Palomar set is greatly increased by the single data point corresponding to 1235+1453 (see \S\ref{dataRedSec}), for which the BAL is dramatically weaker in the Palomar epoch.  If we omit that outlying data point, the scatter of the set with $\Delta t_{sys} < 500$~days decreases significantly to $\sigma \approx 840 \pm 240$~km~s$^{-1}$.  By contrast, removing the largest ``outlier'' from the set with $\Delta t_{sys} > 2000$~days decreases the variance by a much smaller amount, to $1610 \pm 340$~km~s$^{-1}$.  We also investigated the use of the median absolute deviation to estimate the standard deviation $\sigma$ as:
\begin{eqnarray*}
\sigma &\approx& {\rm median}(|x_i - {\rm median}(x_i)|) / 0.6745.
\end{eqnarray*}
The median absolute deviation is less sensitive to outlying points than the sample variance is \citep[e.g.,][]{mmy06}.  In fact, we find $\sigma(\Delta t_{sys}<500$~d$)=880$~km~s$^{-1}$ and $\sigma(\Delta t_{sys}>2000$~d$)=2100$~km~s$^{-1}$, supporting the case that 1235+1453 is an outlier to the trend of increasing spread in $\Delta$EW.  Even so, it is not clear that 1235+1453 is appropriately classified as an outlier; much larger samples will determine the distribution of $\Delta$EW and its apparent time evolution.

It may be the case that the distribution of $\Delta$EW is not Gaussian over times spanning months to years.  For example, sources may be moved to the tail of the $\Delta$EW distribution by strong, temporary ``outbursts'' in continuum emission regions that are not covered by the absorbing outflow.  Such outbursts could be independent of the outflow evolution.  While the data for 1235+1453 suggest that the distribution of $\Delta$EW may be governed by physical processes that make it non-Gaussian, our current sample is not large enough to verify this hypothesis.

\begin{figure} [ht]
  \begin{center}
      \includegraphics[width=5in, angle=270]{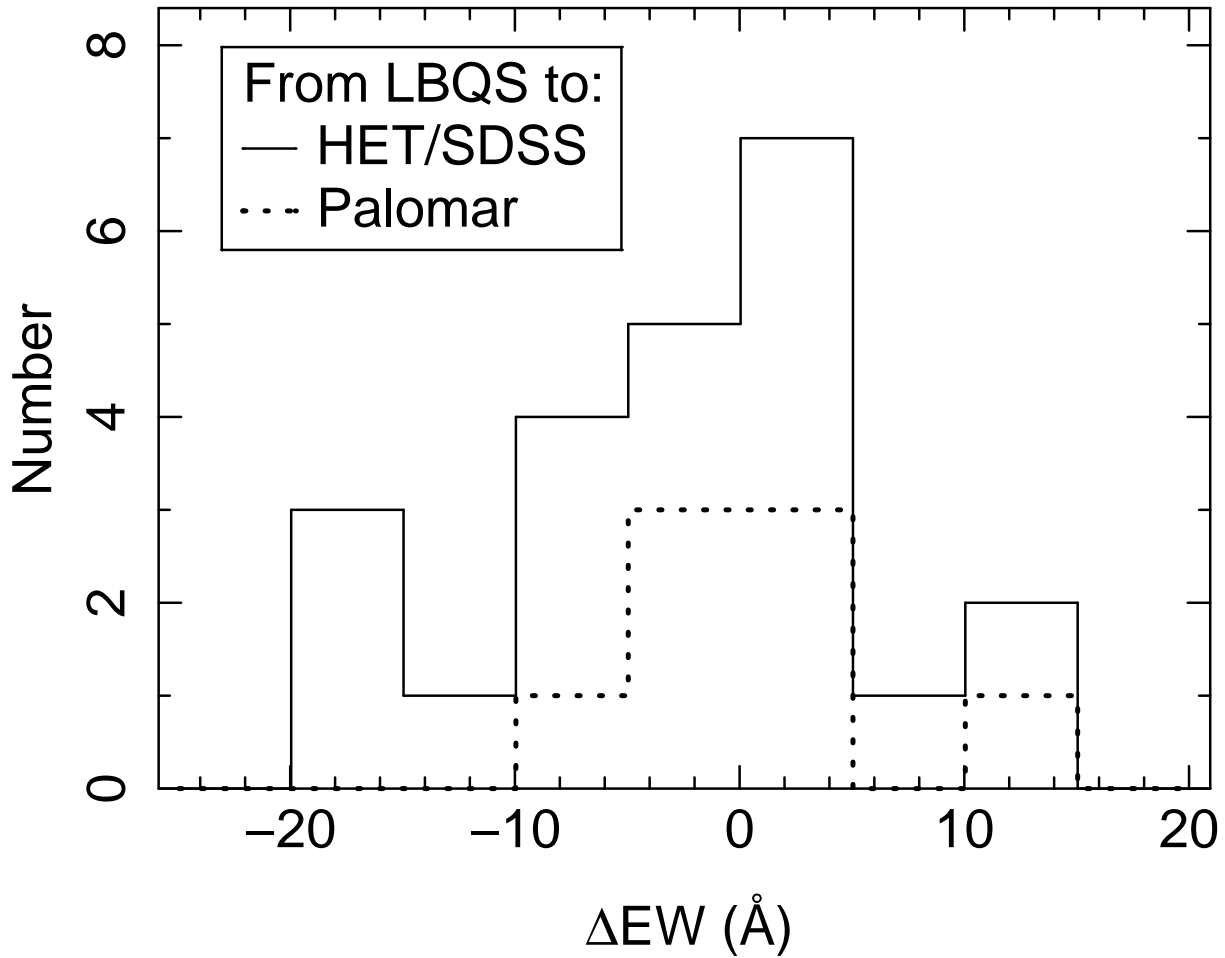}
      \caption{\label{dEWHistFig}Histogram (thin solid line) of $\Delta$EW over 3--7~yr for the combined sample of 23 sources from this work and those of G08 with duplicates removed.  We calculate $\Delta$EW as the difference in EW between the earliest (LBQS) and latest (SDSS or HET) epochs.  Strengthening BALs have $\Delta EW < 0$.  For comparison, we also plot a histogram (dotted line) of $\Delta$EW over time scales of weeks to months between the LBQS and Palomar epochs in our sample, when available.}
   \end{center}
\end{figure}

To analyze the general properties of long-term BAL variability over 3--7 rest-frame years, we combine the sample of this work with the (non-redundant) sources from G08.  In each case, we compute $\Delta$EW, the change in BAL EW between the earliest (LBQS) and latest (either SDSS or HET) epoch.  A histogram of $\Delta$EW values is plotted in Figure~\ref{dEWHistFig}.  For comparison, we have also plotted a histogram of $\Delta$EW on the shorter time scales between LBQS and Palomar epochs for 8~sources using a dotted line in the figure.  The distribution of $\Delta$EW is reasonably symmetric, with 13 of 23 BALs having $\Delta$EW $< 0$, corresponding to strengthening absorption.  While there are subtle differences in source selection between the current work and G08, and a weak bias toward strengthening BALs could be introduced by flux-based quasar detection in the earlier epoch, the overall result indicates no evidence of asymmetry in the increase or decrease of absorption EWs on multi-year time scales.  Asymmetry could occur, for example, if BALs formed very rapidly and then decayed slowly; nearly all identified BALs would then be observed to decay over time.  Any monotonic trends in BAL variation over 3--7 years must either be weaker than we can measure, and/or these trends are dominated by stronger variation on multi-year time scales.

\subsection{Variation as a Function of Absorption Strength\label{fracVarEWSec}}

In this section, we examine whether BAL variation depends on the overall strength (EW) of BAL absorption.  There are several ways to do this.  First, we look for relations between the amount of change in the \ion{C}{4} BAL ($\Delta$EW or $|\Delta EW|$) and a measure of BAL strength (LBQS-epoch EW or $\langle EW \rangle$, defined as the average EW between LBQS and HET epochs).  We find no significant correlations (see Table~\ref{corrTab}) excepting a relation (99.6\% confidence) between $|\Delta EW|$ and the LBQS-epoch EW, and a weaker relation (97\% confidence) between $|\Delta EW|$ and $\langle EW \rangle$.  These relations are not surprising, as stronger BALs are generally broader and have more absorbed regions that can vary.

\begin{figure} [ht]
  \begin{center}
      \includegraphics[width=5in, angle=270]{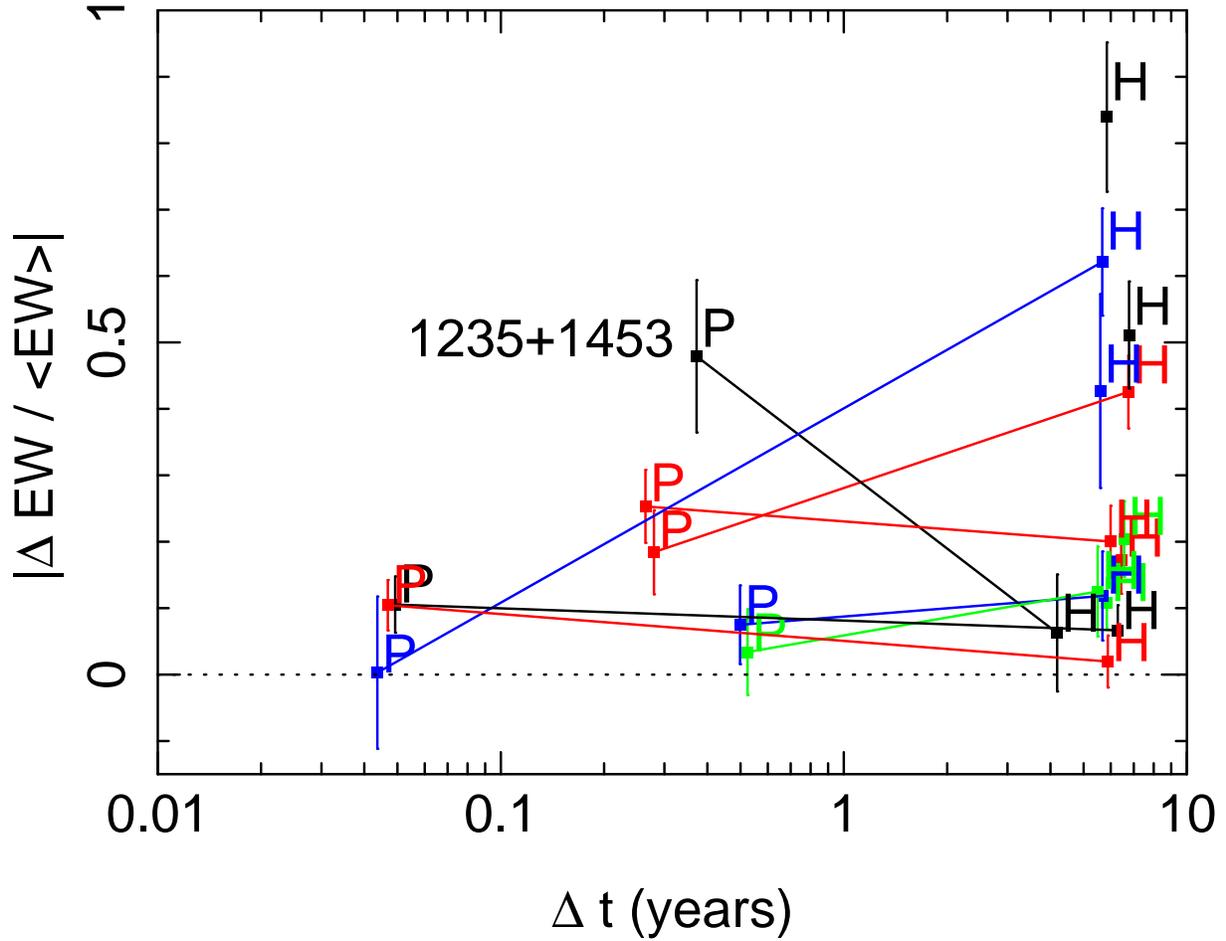}
      \caption{\label{fracEWVarFig}Magnitude of fractional change in EW against the rest-frame time between epochs.  Points marked with ``P'' correspond to Palomar epochs, when available, while points marked with ``H'' correspond to HET epochs.  Solid lines connect Palomar and HET epochs for the same source.    In the color version (online), the points and lines for each source have been given different (arbitrary) colors for visual clarity.  A color version of this figure is available in the online journal.}
   \end{center}
\end{figure}

The amount of time between EW measurements may also affect the degree of change in EW.  Following earlier studies \citep[e.g.,][]{lwbhsyvb07}, we define the fractional change in EW, $|\Delta EW / \langle EW \rangle|$, where the average EW, $\langle EW \rangle$, is calculated by averaging the EW for a given epoch with that of the LBQS epoch.  (Note that we are technically discussing the {\it magnitude} of fractional variation, as we define this quantity to be positive.)  In Figure~\ref{fracEWVarFig}, we plot the magnitude of fractional change in EW as a function of the time interval between observations.  We plot fractional changes in EW for the HET epoch and for the Palomar epoch, when available.  For sources with Palomar spectra, we draw a solid line connecting the Palomar and HET epochs.  As for the measurements of $\Delta$EW, 1235+1453 is once again an outlier; it is the only source with a large decrease in $|\Delta EW / \langle EW \rangle|$ between epochs.

In agreement with previous studies (e.g., \citealt{lwbhsyvb07}; G08), we find that the largest fractional changes are observed on longer time scales.  The cases with the largest fractional change in EW (0009+0219, 0025--0151, and 1314+0116) show reasonably strong BALs with BI$_0 > 1000$~km~s$^{-1}$ in one or both of the LBQS and HET epochs, where BI$_0$ is the traditional ``balnicity index'' \citep{wmfh91} modified to include absorption in the region between $-3000$ and 0~km~s$^{-1}$.  

As previously noted (G08), the large amount of variation observed in some strong BALs could, if continued at the same rate, lead to the disappearance of some BALs over $\sim$15 years.  Because strong BALs seem to have lifetimes longer than that (\S\ref{evolAbsSec}), we would expect that the fractional EW variation ``turns over'' on longer time scales for these highly-variable sources and that they do not vary monotonically.  We also note that 2 of the 3 sources with $|\Delta EW / \langle EW \rangle| > 0.5$ are actually increasing in strength, rather than decreasing.

\subsection{Variation on Sub-Year and Multi-Year Time Scales\label{varTimeScaleSec}}

We also test for indications that changes in a significant number of {\it individual} BAL EWs may be monotonic over months to years.  To do this, we compute $\Delta$EW$_{PL}$, defined as the difference in EW between the Palomar and LBQS epochs, and $\Delta$EW$_{HP}$, defined as the difference between HET and Palomar epochs.  The typical rest-frame time between the LBQS and Palomar epochs is a few weeks to months, while the time between Palomar and HET epochs is 3.8--6.5~yr.  We find no evidence of a correlation between $\Delta$EW$_{PL}$ and $\Delta$EW$_{HP}$ for the 8 sources in our sample that have spectra in all three epochs.  While some BALs may vary monotonically over long time scales (including sources such as 1314+0116 in our sample), the variations observed in a BAL over a few months do not allow us to reliably predict the longer-term variation over several years for BALs in general.

\begin{figure} [ht]
  \begin{center}
      \includegraphics[width=5in, angle=270]{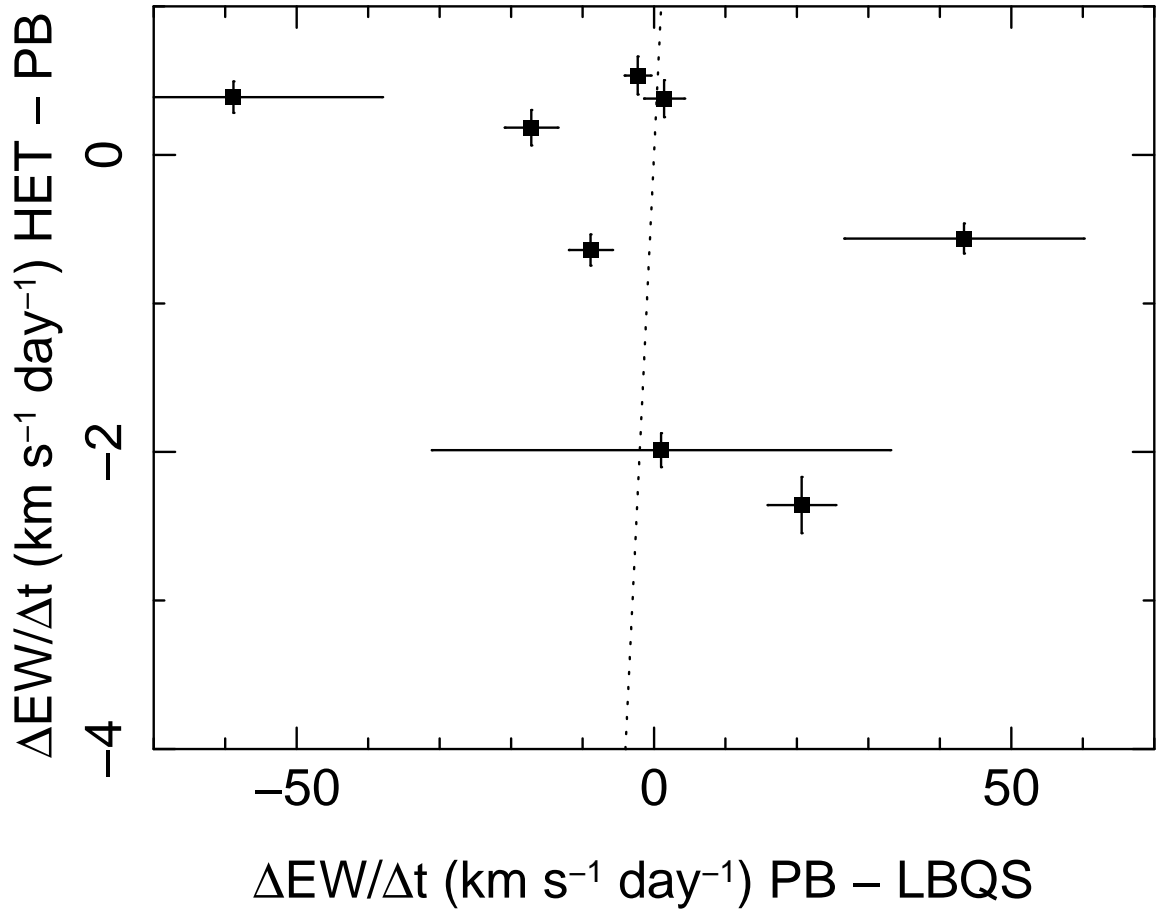}
      \caption{\label{dEWdtFig}Comparison of rest-frame $\Delta EW / \Delta t$ measured for the LBQS-to-Palomar epochs ($x$-axis) and the Palomar-to-HET epochs ($y$-axis).  The time duration between the Palomar and HET epochs is much longer, leading to a much smaller range of values on the $y$-axis and larger error bars in the $x$-direction.  The dotted line indicates the line $x = y$.}
   \end{center}
\end{figure}

The rate of change in EW, $\Delta EW/ \Delta t$, is also unrelated on shorter (LBQS to Palomar epochs) and longer (Palomar to HET epochs) time scales.  We find no significant correlation in $\Delta EW/ \Delta t$ measured for these two time scales.  In fact, several BALs show opposite signs for $\Delta EW / \Delta t$, as the absorption weakened between one set of observations and strengthened in the other (Figure~\ref{dEWdtFig}).

\subsection{Evolution of BAL Absorption\label{evolAbsSec}}

In G08, we placed a loose lower limit on the lifetime of strong BALs of $t_{BAL} > 18$~yr, based on the observation that the strong BALs in the G08 study had not disappeared over typical rest-frame times of $\sim$4.3~yr.  G08 defined ``strong BALs'' to have BI$_0 > 100$~km~s$^{-1}$.  All but one of the BALs in our current study have BI$_0 > 0$ for both the LBQS and HET epochs.  The exception is 0022+0150, for which the absorption has weakened and BI$_0$ is measured to be zero in the HET epoch.  In fact, the BI$_0$ measurement is extremely sensitive to continuum placement, and a small adjustment to the continuum would give a measurement of BI$_0 > 0$~km~s$^{-1}$.

We model the probability of observing a BAL to disappear to be $\Delta t_{sys} / t_{BAL}$, where $\Delta t_{sys}$ is the rest-frame time between observing epochs and $t_{BAL}$ is the typical BAL lifetime.  Given that we have 19 strong BALs (with BI$_0 > 100$~km~s$^{-1}$ in at least one epoch) in our sample (including 5 non-redundant sources from G08) and that the median $\Delta t_{sys}$ is 5.8~yr for this augmented sample, we now find that we would have a $\ge$90\% chance of observing more than one case (0022+0150) in which a BAL disappears if $t_{BAL} \la 30$~yr in the rest frame.  (The lifetime rises to $\gtrsim$50~yr if we allow 0022+0150 to be classified as a BAL in the HET epoch.)

While BALs may be transient in individual cases, and weaker BALs (and mini-BALs) are not included in this estimate, it seems that strong BALs are overall rather stable on human time scales.  BAL outflows therefore appear to travel large distances; material moving constantly at 10,000~km~s$^{-1}$ will traverse $\sim$0.3--0.5~pc in 30--50~yr.  Even if we are seeing a ``standing outflow pattern'' in which new material continually enters at an angle to our line of sight \citep[e.g.,][]{d97}, the material that has left our line of sight can still reach large distances over the BAL trough lifetime unless it is decelerated or obstructed after it has moved out of view.  If BAL outflows are launched at accretion disk radii much smaller than 1~pc, they may still interact with any parsec-scale structures such as an obscuring torus (if present in luminous quasars at outflow inclination angles).

The absorption for 0022+0150 has weakened to the point where the absorption trough widths are becoming narrower than the $2000$~km~s$^{-1}$ cutoff required for BAL absorption.  0022+0150 may be more properly classified as a ``mini-BAL'' QSO in the HET data and would be another demonstration of the link between BAL and mini-BAL QSOs \citep[e.g.,][]{akdjb99, gbcg02, gbgs09}.  \citet{lhcg09} have recently pointed out evolution in the opposite direction for the nearby Seyfert~1 galaxy WPVS~007 (at $z \approx 0.03$).  In that source, BAL troughs emerged to augment the mini-BALs that were seen in an earlier epoch.

\subsection{Comparing Variation in \ion{Si}{4} and \ion{C}{4} BAL Shape\label{siIVCIVShapeVarSec}}

Detailed comparisons of absorption profiles among ionization species can help to map the ionization structure of the BAL outflow.  For example, \citet{akdjb99} observed that the ionization-dependent structure of BAL components in PG~$0946+301$ was best explained if BAL outflows were structured to have lower-ionization ``tubes'' of material embedded in or associated with higher-ionization flows.  The lower-ionization material could have a different (and smaller overall) covering factor than the higher-ionization material, leading to somewhat different trough shapes.  As a consequence of the outflow geometry, lower-ionization material is also seen in a narrower velocity range.  In fact, \ion{Si}{4} troughs do appear to be narrower than \ion{C}{4} troughs for BAL QSOs in general \citep[e.g.,][]{gjbhswasvgfy09}.

Variability adds a new dimension to this picture.  If absorption depths are determined by the fraction of the emitter covered by the outflow, then synchronous variation in absorption from different ions would require that the flow be structured so that different ionization components can vary in unison.  On the other hand, if the absorption is not saturated, changes in the outflow ionization level may lead to more complex changes in the optical depth of the outflow.  In this section, we discuss the relation between \ion{Si}{4}~$\lambda$1400 and \ion{C}{4}~$\lambda$1549 BAL troughs in different epochs in order to determine whether they vary synchronously.  \ion{Si}{4} and \ion{C}{4} have ionization potentials of 45.1 and 64.5~eV, respectively, and both ions have large ionization fractions over a range of (ionizing-continuum-dependent) absorber ionization levels.

We use a straightforward approach in which we directly compare the observed changes in BAL absorption for \ion{Si}{4} and \ion{C}{4}.  We do this using the four ratio spectra --- LBQS and HET epochs for \ion{Si}{4} and \ion{C}{4} --- that are obtained when the observed spectra are divided by our emission model.  We smooth each spectrum using a boxcar filter of width 3 bins and then rebin onto a common velocity grid.  Our velocity grid only extends out to $-20,000$~km~s$^{-1}$ to avoid contamination by unmodeled features blueward of \ion{Si}{4}.  We then subtract the ratio spectrum of the earlier (LBQS) epoch from that of the later (HET) epoch.  This gives us two ``difference spectra,'' which we call $\Delta S(v)$ for the \ion{Si}{4} absorption region and $\Delta C(v)$ for \ion{C}{4} absorption.  $\Delta S(v)$ represents the change in the \ion{Si}{4} absorption profile, and $\Delta C(v)$ does the same for \ion{C}{4}.  Note that $\Delta S(v)$ and $\Delta C(v)$ also include information about the variation of narrower absorption features that may reside outside the BAL trough.

\begin{figure} [ht]
  \begin{center}
      \includegraphics[width=3.0in, angle=270]{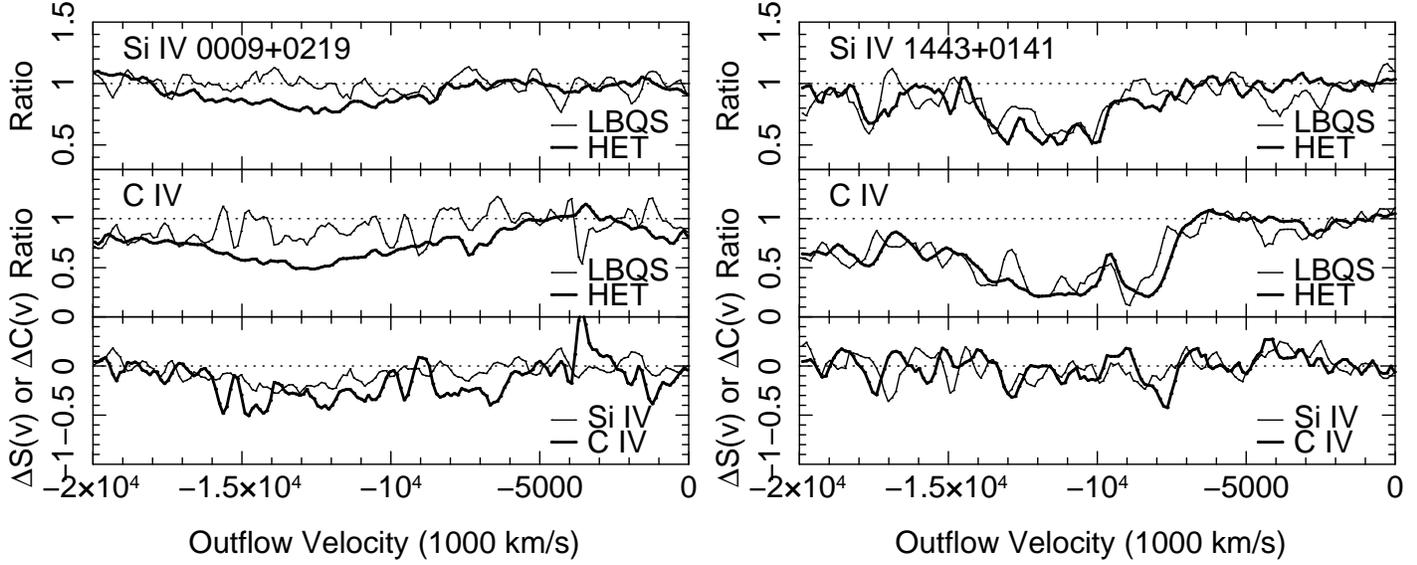}
      \caption{\label{diffSpecsSiIVCIVCasesFig}For our two candidate cases for coordinated \ion{Si}{4} and \ion{C}{4} absorption variation (0009+0219, on left, and 1443+0141, on right), we show the ratio spectra (with emission model divided out) for \ion{Si}{4} (top panel) and \ion{C}{4} (middle panel) absorption regions out to $-20000$~km~s$^{-1}$.  LBQS epochs are shown with thin lines and HET epochs are shown with thick lines in those two panels.  In the bottom panel, we plot $\Delta S(v)$ as a thin line and $\Delta C(v)$ as a thick line.  These ``difference spectra,'' described in \S\ref{siIVCIVShapeVarSec}, represent the difference between the absorber transmission for \ion{Si}{4} and \ion{C}{4}, respectively, as a function of velocity.  Related changes in the absorber strength can be seen in at least a subset of velocity regions for each ion (e.g., increased absorption at --8000 and --13,000~km~s$^{-1}$ and decreased absorption at --4000~km~s$^{-1}$ for 1443+0141).}
   \end{center}
\end{figure}

Although the \ion{Si}{4} and \ion{C}{4} line doublets have different widths that affect the different BAL trough shapes, this rough comparison still identifies correlations at $>$99.9\% significance between $\Delta S(v)$ and $\Delta C(v)$ for two cases, 0009+0219 and 1443+0141, according to a Spearman rank correlation test.  Several other spectra show evidence for correlation at $>95$\% significance levels, although visual inspection indicates that the correlations for these may be influenced by a relatively small number of points on the boundaries of the $\Delta S(v)$ vs. $\Delta C(v)$ relation.  At least 0009+0219 and 1443+0141 show coordinated changes between epochs over a range of velocities.  Figure~\ref{diffSpecsSiIVCIVCasesFig} shows $\Delta S(v)$ and $\Delta C(v)$ for these sources.  For a few other sources, the changes are suggestive or are strongly related over a smaller subset of the full velocity range.

\subsection{Comparing Variation in \ion{Si}{4} and \ion{C}{4} BAL Strength\label{siIVCIVStrengthVarSec}}

In this section, we are not concerned with the variation of absorption trough shapes, but only with the total strength of absorption.  The absorption EW differs from ion to ion in a single epoch, with \ion{Si}{4} absorption generally being weaker than \ion{C}{4} absorption \citep[e.g.,][]{gjbhswasvgfy09}.  This could be due to a number of factors such as ionization levels, elemental abundances, and BAL outflow geometry having an ion-dependent covering factor \citep[e.g.,][]{akdjb99}.

\begin{figure} [ht]
  \begin{center}
      \includegraphics[width=5in, angle=270]{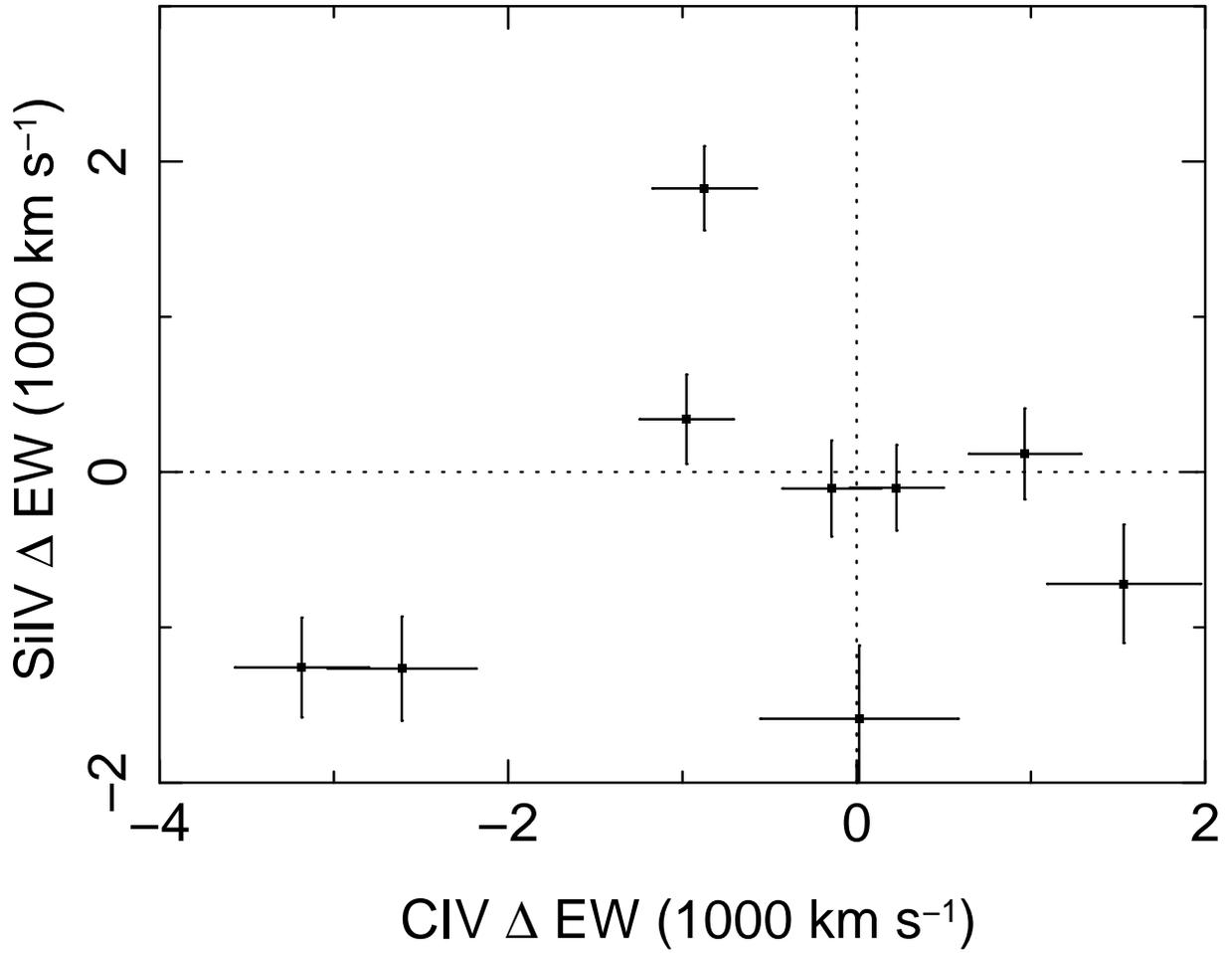}
      \caption{\label{compareSidEWsVsCIVFig}Change in EW ($\Delta$EW) between LBQS and HET epochs for \ion{Si}{4} and \ion{C}{4} absorption regions for sources at sufficiently high redshift to have full HET coverage of the \ion{Si}{4} absorption region.  The \ion{Si}{4} and \ion{C}{4} EWs are integrated out to $-20,000$~km~s$^{-1}$.}
   \end{center}
\end{figure}

In Figure~\ref{compareSidEWsVsCIVFig}, we plot $\Delta$EW between LBQS and HET epochs for \ion{Si}{4} absorption regions compared to $\Delta$EW for \ion{C}{4} absorption regions.  Here again, we only integrate out to $-20,000$~km~s$^{-1}$ for both ions in order to avoid emission model uncertainty for \ion{Si}{4} at higher outflow speeds.  We only include the 9~sources at sufficiently high redshift to have HET coverage out to $-20,000$~km~s$^{-1}$ for \ion{Si}{4}.  Our calculated BAL EW values are listed in Table~\ref{siCIVEW20Tab}.  We find no significant correlation of $\Delta$EW between \ion{Si}{4} and \ion{C}{4}.  We also find no significant correlation of $\langle EW \rangle$ between the two ions.

\begin{figure} [ht]
  \begin{center}
      \includegraphics[width=5in, angle=270]{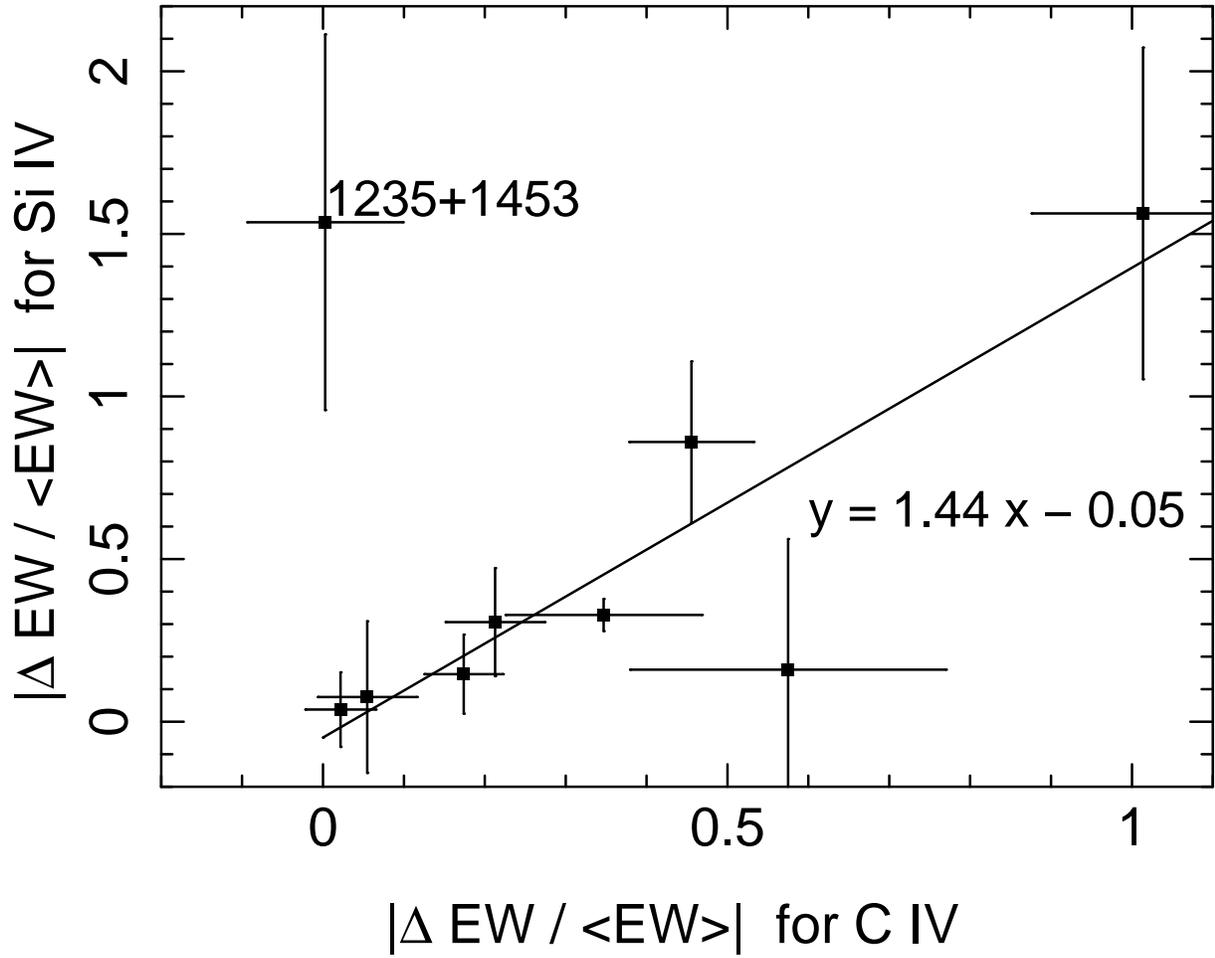}
      \caption{\label{compareBALVarSiFracdEWsVsCIVFig}Same as Figure~\ref{compareSidEWsVsCIVFig} but showing the fractional change in EW for \ion{Si}{4} ($y$-axis) and \ion{C}{4} ($x$-axis).  The solid line represents a fit to the data with the outlier 1235+1453 removed.}
   \end{center}
\end{figure}

There is a suggestion of a correlation in the fractional change in EW, $|\Delta EW / \langle EW \rangle|$, as shown in Figure~\ref{compareBALVarSiFracdEWsVsCIVFig}.  The correlation for the full set of sources is not highly significant, but a stronger correlation ($>99$\% confidence) would be observed if 1235+1453, the outlier at the top left of the plot, were removed.  As previously discussed (\S\ref{dataRedSec}), this source may have contaminating ``emission'' that artificially increases $\Delta$EW and decreases $\langle EW \rangle$, moving the data point for 1235+1453 upward in Figure~\ref{compareBALVarSiFracdEWsVsCIVFig}.  We use the IDL FITEXY code to fit $F({\rm Si IV})$ against $F({\rm C IV})$, where $F(x)$ is the fractional change in EW for ion $x$, accounting for the formally-derived errors.  We find a fit of:
\begin{eqnarray}
F({\rm Si~IV}) &=& (1.44 \pm 0.45) F({\rm C~IV}) + (-0.05 \pm 0.19)\label{fracEWChangeFitEqn}
\end{eqnarray}
for the sample with 1235+1453 omitted.  The fit is shown as a solid line in Figure~\ref{compareBALVarSiFracdEWsVsCIVFig}.

Correlations in $|\Delta EW / \langle EW \rangle|$ can arise, for example, when changes in the ionization level or geometry affect the same fraction of material for each ion.  While the ratios of \ion{Si}{4} to \ion{C}{4} may vary among different QSO outflows, it is plausible that the fraction of material that changes will be the same for both ions in a given outflow.  For models of such scenarios, we would not necessarily expect to see a correlation between $\Delta$EW for the two ions, as $\Delta$EW can depend on the amount of absorbing material that is present along the line of sight for each ion (and also the precise mode of variation).

Under the assumption that 1235+1453 can be classified as an outlier, our results would indicate that \ion{Si}{4} absorption varies in concert with \ion{C}{4} absorption, and in at least some cases, the changes in velocity-dependent absorption profiles can be highly related (\S\ref{siIVCIVShapeVarSec}).  However, according to Figure~\ref{compareSidEWsVsCIVFig}, it does appear that overall \ion{Si}{4} absorption can strengthen while \ion{C}{4} weakens, and {\it vice versa}.  If variation were purely due to changes in the emission covering factor, we would expect that both ions would always strengthen or weaken together.

\subsection{Comparison With Emission Line Variation\label{compareEmVarSec}}

The strength (EW) of \ion{C}{4} line emission is known to be related to the UV continuum luminosity \citep[e.g.,][]{b77}.  This ``global Baldwin effect'' relation apparently extends to higher-energy continuum emission as well, with the \ion{C}{4} emission EW being anti-correlated to the continuum monochromatic luminosity at 2500~\AA\ and correlated to the monochromatic luminosity at 2~keV \citep{g98, wlz98, gbs08}.  Additionally, an ``intrinsic'' Baldwin effect is observed in which \ion{C}{4} EW variation is anti-correlated with continuum luminosity in a single object as its spectrum varies over time \citep[e.g.,][]{krk90}.  While BAL QSOs have weaker \ion{C}{4} line emission in general than non-BAL QSOs do, \ion{C}{4} emission in BAL QSOs still follows a ``Baldwin effect'' type trend with UV luminosity \citep{gjbhswasvgfy09}.

Motivated by these considerations, we have briefly investigated whether we can use \ion{C}{4} emission EW as a proxy for the ionizing continuum luminosity to search for evidence of photoionization-driven BAL variation.  We find no significant correlations between emission line EW (listed in Table~\ref{cIVEmEWTab}) and the \ion{C}{4} absorption EW.  We also find no correlation between the {\it changes} in \ion{C}{4} emission EW and \ion{C}{4} BAL absorption.  We do not draw strong physical conclusions from this result because any underlying correlation could be weakened by factors such as different ionization time scales for the emitting and absorbing material as well as the fact that the observed \ion{C}{4} emission is integrated over a large spatial region.  Furthermore, we are unable to account for any variable absorption that would have a velocity profile broad and smooth enough to affect the entire broad emission line without leaving obvious absorption signatures.

\subsection{The Stability of BAL Outflows\label{nonVarSec}}

While BAL shapes differ greatly from one QSO to another, numerous cases in the current study and also G08 demonstrate that the factors shaping BAL outflows (such as the velocity-dependent covering factor) remain relatively stable or even unchanged for years over much of the outflow velocity range.  As previous researchers have noted \citep[e.g.,][]{d97}, BAL outflow models are challenged to explain both the range of variation effects seen in the spectra and also the remarkable stability in many velocity components of BAL outflows.  In some cases, the velocity structure may be stabilized by physical effects associated with the acceleration mechanism, such as line-locking \citep[e.g.,][]{gbwccl04}.

Over a median observed time of 5.8~yr in the rest frame, material flowing outward at 10,000~km~s$^{-1}$ travels a radial distance of $2\times 10^{17}$~cm, which is $12,400$~$GM_{BH}/c^2$ for a central black hole of mass $M_{BH} = 10^8$~M$_{\astrosun}$.  If the BAL outflow is associated with a rotating accretion disk and resides at a radius of $10^{17}$~cm, the Keplerian frequency of circular rotation is $\omega = \sqrt{G M_{BH} / r^3} = 0.11$~rad~yr$^{-1}$.  This corresponds to a velocity of $3600$~km~s$^{-1}$ perpendicular to the line of sight, or a distance of $7\times 10^{16}$~cm covered over 5.8 years.  Over a typical BAL lifetime of presumably $\gtrsim$30--50~yr (\S\ref{evolAbsSec}), the azimuthal distances traveled would be $\gtrsim$5--9 times larger, or $\gtrsim 0.1-0.2$~pc.

If the BAL outflow is launched from a rotating accretion disk, it must be possible to maintain relatively stable outflow structures over such large time and size scales.  This is true whether we are observing a ``standing outflow pattern'' (in which the outflow material is continually replenished) or the same absorbing material over time.  In the latter case, the outflow structure must also be stable as the absorber travels large radial distances.  For a quasar with $M_{BH} = 10^8$~M$_{\astrosun}$, the broad emission line region radius (estimated from reverberation measurements of Balmer lines) is $\sim 1.7\times 10^{17}$~cm \citep{ksnmjg00}, and higher-velocity material in the BAL outflow has presumably traveled beyond this distance between the LBQS and HET epochs.

\subsection{BAL Variation Patterns\label{varPatternsSec}}

In this work, we have used absorption measures integrated over the entire absorption region (e.g., the absorption EW) to describe the variation of BAL troughs.  In an earlier work (G08), we developed metrics that could be used to characterize the properties of variation in narrower velocity regions of a single BAL.  We do not apply those metrics in the current study because they require SDSS-quality spectra with higher resolution and S/N in at least one epoch in order to be effective.  However, visual inspection indicates that the variation in our sample is generally consistent with the G08 metrics.  The variable regions of a BAL are typically narrower subsets of the full absorption trough, and BALs can vary in some absorption regions while remaining relatively constant in others (e.g., 0019+0107 at 1430--1480~\AA, Figure~5b; 0021--0213 at 1460--1480~\AA, Figure~5c).  Variation occurs at both lower (e.g., 0025--0151 at 1510--1530~\AA, Figure~5e) and higher outflow velocities (e.g., 1314+0116 at 1420--1480~\AA\ in HET, Figure~5l; 1442--0011 at 1430--1450~\AA\ in HET, Figure~5m).  We see variation in shallower (e.g., 0019+0107, Figure~5b) as well as deeper (e.g., 0021--0213, Figure~5c; 1231+1320 at 1520~\AA, Figure~5g) troughs, and the amount of change in the absorption depth is commonly $\lesssim 30$\% of the continuum.

Cross-calibration of the wavelengths between the LBQS and HET epochs reveals two potentially interesting cases of changes in velocity structure.  In the first case, 0022+0150 (Figure~5d), there are three relatively weak (by BAL standards) absorption features at $\sim$1430, 1490, and 1520~\AA.  The strongest feature, at $\sim$1520~\AA, has weakened over time.  Additionally, there is an absorption feature on the blue side of the Ly$\alpha$ line (Figure~\ref{specAndModelFig}).  There is no clear way to line up the emission and absorption features between the LBQS and HET epochs so these features match.  If we line up the spectra so that the red side of Ly$\alpha$ emission matches, then the absorption features have shifted $5-10$~\AA\ blueward.  Alternatively, we can match up most of the broader and narrower absorption features, but the red wing of Ly$\alpha$ has moved redward and the weak BAL absorption at 1430~\AA\ has moved blueward by about 5~\AA\ over time.  Because the cross-calibration is anchored by absorption features within 130~\AA\ on both the blue and red sides of the 1430~\AA\ feature, any wavelength calibration issues would have to be highly nonlinear to account for the shift of the 1430~\AA\ feature.

The second candidate for absorption acceleration is 1240+1607 (Figure~5j).  While narrower absorption features match up well at $\sim$1440, 1525, and 1535~\AA, the broader trough from $\sim 1490-1510$~\AA\ has shifted blueward by about 5~\AA\ over time.  In this case, the absorption in the Palomar epoch from $1510-1540$~\AA\ also follows a somewhat different pattern than the LBQS and HET epochs, suggesting that the absorption variation in this source may be particularly complex.

These examples demonstrate issues that complicate a definitive search for BAL acceleration.  If the trough components have truly accelerated in these cases, the change in velocity is $\sim$1000~km~s$^{-1}$ over 5.6~yr (0022+0150) and 6.0~yr (1240+1607) for these two sources.  The corresponding acceleration is a factor of $\sim$4--18 larger than that observed in other sources \citep{vi01, rvs02, hsher07}.  Of course, we cannot rule out the possibility that the troughs have simply changed shape rather than accelerating.  If absorption deepens on the previously-unabsorbed blue side and weakens on the red side of a trough, the absorption feature will appear to accelerate.

\section{SUMMARY AND CONCLUSIONS\label{concSec}}

Here, we briefly summarize results in our study.

\begin{enumerate}
\item{Multi-year BAL variability is complex; there is no clear pattern of \ion{C}{4} absorption variation over \mbox{5--7~yr} in our sample of 14 BAL QSOs at $z > 2.1$.}
\item{The scatter in $\Delta$EW increases with time to $\sim$1800~km~s$^{-1}$ over 5.8~yr.  We find no evidence for asymmetry in the distribution of $\Delta$EW about 0~km~s$^{-1}$ that would indicate that BALs strengthen and weaken on dramatically different time scales.}
\item{In one source, 1235+1453, we find evidence for a variable, blue continuum component that is relatively unabsorbed by the BAL outflow.}
\item{Although individual BALs may strengthen or weaken monotonically, BALs do not generally evolve monotonically over 3--7~yr.  Variation on multi-month time scales does not predict variation over years in an obvious way.}
\item{Based on the fact that we saw at most 1~BAL (in 0022+0150) formally leaving our augmented sample of 19~strong BALs, we estimate that at least stronger BAL features have a lifetime $\gtrsim$30~yr.  The absorption in 0022+0150 did not fully disappear; instead, the absorption weakened to mini-BAL status.}
\item{The evolution of substructures in \ion{Si}{4} and \ion{C}{4} BAL trough shapes appears to be related in at least two cases in our sample.}
\item{While $\Delta$EW is not strongly correlated over 5.8~yr for \ion{Si}{4} and \ion{C}{4} BAL trough evolution, there does appear to be a strong correlation in the fractional change in EW for \ion{Si}{4} and \ion{C}{4} troughs.  This result depends on the removal of one outlier, 1235+1453, which (as we discuss) has unusual properties in several respects.}
\item{Given their apparently long lifetimes and high velocities, BAL outflows can cover large radial and azimuthal distances relative to the accretion disk center.  The stability of many BAL troughs (in whole or in specific velocity ranges) over years is an important consideration for models in which BAL outflows are associated with rotating accretion disks.}
\item{We cross-calibrated the wavelengths of spectra between epochs using narrower absorption and emission features and identified two cases (0022+0150 and 1240+1607) of \ion{C}{4} BALs that showed evidence for possible changes in velocity structure of a magnitude $\sim$1000~km~s$^{-1}$ over $\approx$6~yr.  Follow-up is needed for these and other candidate sources to determine the nature of possible absorber acceleration.}
\end{enumerate}

The current work and previous studies of BAL variability in QSO samples (e.g., \citealt{b93, lwbhsyvb07}; G08) demonstrate that there is still much to learn about the structure, evolution, and influence of BAL outflows, as well as the physics that launches and shapes them.  Improved samples of BAL spectra are clearly needed to examine the nature of BAL variation and acceleration, the ionization structure and evolution of BAL outflows, links with other absorption phenomena such as mini-BALs, and the long-term evolution of BAL outflows.

Given the practical issues involved with observing UV spectra over long time frames, we suggest some approaches that would be particularly advantageous for the ongoing study of high-velocity QSO outflows.  Firstly, larger samples are needed.  The SDSS archive will provide an excellent baseline for studies of BALs at higher resolution and S/N, and the SDSS-III surveys will provide an additional epoch of flux-calibrated spectroscopy for many SDSS BAL QSOs.

Secondly, intensive monitoring of a set of variable BALs, ideally at higher resolution, could resolve interesting questions about BAL structure and how BALs evolve over shorter time scales.  Among other things, spectra from such a campaign could reveal substructures in the outflow that vary in concert (in the BAL of a single ion or from multiple ions), and would also enable a search for rapid photoionization-driven changes in the outflow.  Candidates for BAL acceleration, such as 0022+0150, 1240+1607, and SDSS~J$024221.87+004912.6$ \citep{hsher07} should also be re-observed over months to years to test whether acceleration has continued.

Ideally, we would like to determine a ``structure function for BAL variation'' that would be a useful input to models of quasar outflows.  The computation of structure functions and other detailed variability metrics will become possible as large-scale spectroscopic surveys such as SDSS-III re-observe the large samples of BAL quasars identified in earlier SDSS surveys.

\acknowledgements

We gratefully acknowledge support from NASA Chandra grant AR9-0015X (R.R.G.), NSF grant AST07-09394 (R.R.G.), NASA LTSA grant NAG5-13035 (W.N.B., D.P.S.), NSF grant AST06-07634 (D.P.S.), and the National Science and Engineering Research Council of Canada (S.C.G.).

Funding for the SDSS and SDSS-II has been provided by the Alfred P. Sloan Foundation, the Participating Institutions, the National Science Foundation, the U.S. Department of Energy, the National Aeronautics and Space Administration, the Japanese Monbukagakusho, the Max Planck Society, and the Higher Education Funding Council for England.  The SDSS Web site \hbox{is {\tt http://www.sdss.org/}.}

The Hobby-Eberly Telescope is a joint project of the University of Texas at Austin, the Pennsylvania State University, Stanford University, Ludwig-Maximilians-Universit\"{a}t M\"{u}nchen, and Georg-August-Universit\"{a}t G\"{o}ttingen. The HET is named in honor of its principal benefactors, William P. Hobby and Robert E. Eberly.

The Marcario Low Resolution Spectrograph is named for Mike Marcario of High Lonesome Optics who fabricated several optics for the instrument but died before its completion. The LRS is a joint project of the Hobby-Eberly Telescope partnership and the Instituto de Astronom\'{i}a de la Universidad Nacional Aut\'{o}noma de M\'{e}xico.


\bibliographystyle{apj3}
\bibliography{apj-jour,bibliography}

\begin{thebibliography}{49}
\expandafter\ifx\csname natexlab\endcsname\relax\def\natexlab#1{#1}\fi
\expandafter\ifx\csname url\endcsname\relax
  \def\url#1{{\tt #1}}\fi
\expandafter\ifx\csname urlprefix\endcsname\relax\def\urlprefix{URL }\fi
\providecommand{\eprint}[2][]{\url{#2}}

\bibitem[\protect\astroncite{{Adelman-McCarthy} et~al.}{2007}]{a-m+07}
{Adelman-McCarthy}, J.~K., et~al. 2007, \apjs, 172, 634

\bibitem[\protect\astroncite{{Arav} et~al.}{1999{\natexlab{a}}}]{ablgwbd99}
{Arav}, N., {Becker}, R.~H., {Laurent-Muehleisen}, S.~A., {Gregg}, M.~D.,
  {White}, R.~L., {Brotherton}, M.~S., \& {de Kool}, M. 1999{\natexlab{a}},
  \apj, 524, 566

\bibitem[\protect\astroncite{{Arav} et~al.}{1999{\natexlab{b}}}]{akdjb99}
{Arav}, N., {Korista}, K.~T., {de Kool}, M., {Junkkarinen}, V.~T., \&
  {Begelman}, M.~C. 1999{\natexlab{b}}, \apj, 516, 27

\bibitem[\protect\astroncite{{Baldwin}}{1977}]{b77}
{Baldwin}, J.~A. 1977, \apj, 214, 679

\bibitem[\protect\astroncite{{Barlow}}{1993}]{b93}
{Barlow}, T.~A. 1993, Ph.D. thesis, Univ. California, San Diego

\bibitem[\protect\astroncite{{Barlow} et~al.}{1989}]{bjb89}
{Barlow}, T.~A., {Junkkarinen}, V.~T., \& {Burbidge}, E.~M. 1989, \apj, 347,
  674

\bibitem[\protect\astroncite{{Barlow} et~al.}{1992}]{bjbwmk92}
{Barlow}, T.~A., {Junkkarinen}, V.~T., {Burbidge}, E.~M., {Weymann}, R.~J.,
  {Morris}, S.~L., \& {Korista}, K.~T. 1992, \apj, 397, 81

\bibitem[\protect\astroncite{{Dai} et~al.}{2008}]{dss08}
{Dai}, X., {Shankar}, F., \& {Sivakoff}, G.~R. 2008, \apj, 672, 108,
  \eprint{arXiv:0704.2882}

\bibitem[\protect\astroncite{{de Kool}}{1997}]{d97}
{de Kool}, M. 1997, in Mass Ejection from Active Galactic Nuclei, eds.
  N.~{Arav}, I.~{Shlosman}, \& R.~J. {Weymann}, vol. 128 of {\em Astronomical
  Society of the Pacific Conference Series\/}, 233--+

\bibitem[\protect\astroncite{{Edelson} et~al.}{2002}]{etpvmmdw02}
{Edelson}, R., {Turner}, T.~J., {Pounds}, K., {Vaughan}, S., {Markowitz}, A.,
  {Marshall}, H., {Dobbie}, P., \& {Warwick}, R. 2002, \apj, 568, 610

\bibitem[\protect\astroncite{{Foltz} et~al.}{1987}]{fchmtwa87}
{Foltz}, C.~B., {Chaffee}, F.~H., Jr., {Hewett}, P.~C., {MacAlpine}, G.~M.,
  {Turnshek}, D.~A., {Weymann}, R.~J., \& {Anderson}, S.~F. 1987, \aj, 94, 1423

\bibitem[\protect\astroncite{{Gallagher} et~al.}{2002}]{gbcg02}
{Gallagher}, S.~C., {Brandt}, W.~N., {Chartas}, G., \& {Garmire}, G.~P. 2002,
  \apj, 567, 37

\bibitem[\protect\astroncite{{Gallagher} et~al.}{2006}]{gbcpgs06}
{Gallagher}, S.~C., {Brandt}, W.~N., {Chartas}, G., {Priddey}, R., {Garmire},
  G.~P., \& {Sambruna}, R.~M. 2006, \apj, 644, 709

\bibitem[\protect\astroncite{{Gallagher} et~al.}{2004}]{gbwccl04}
{Gallagher}, S.~C., {Brandt}, W.~N., {Wills}, B.~J., {Charlton}, J.~C.,
  {Chartas}, G., \& {Laor}, A. 2004, \apj, 603, 425

\bibitem[\protect\astroncite{{Ganguly} \& {Brotherton}}{2008}]{gb08}
{Ganguly}, R. \& {Brotherton}, M.~S. 2008, \apj, 672, 102

\bibitem[\protect\astroncite{{Gibson} et~al.}{2009{\natexlab{a}}}]{gbgs09}
{Gibson}, R.~R., {Brandt}, W.~N., {Gallagher}, S.~C., \& {Schneider}, D.~P.
  2009{\natexlab{a}}, \apj, 696, 924

\bibitem[\protect\astroncite{{Gibson} et~al.}{2008{\natexlab{a}}}]{gbs08}
{Gibson}, R.~R., {Brandt}, W.~N., \& {Schneider}, D.~P. 2008{\natexlab{a}},
  \apj, 685, 773

\bibitem[\protect\astroncite{{Gibson} et~al.}{2008{\natexlab{b}}}]{gbsg08G08}
{Gibson}, R.~R., {Brandt}, W.~N., {Schneider}, D.~P., \& {Gallagher}, S.~C.
  2008{\natexlab{b}}, \apj, 675, 985 (G08)

\bibitem[\protect\astroncite{{Gibson}
  et~al.}{2009{\natexlab{b}}}]{gjbhswasvgfy09}
{Gibson}, R.~R., et~al. 2009{\natexlab{b}}, \apj, 692, 758

\bibitem[\protect\astroncite{{Green}}{1998}]{g98}
{Green}, P.~J. 1998, \apj, 498, 170

\bibitem[\protect\astroncite{{Gregg} et~al.}{2006}]{gbd06}
{Gregg}, M.~D., {Becker}, R.~H., \& {de Vries}, W. 2006, \apj, 641, 210

\bibitem[\protect\astroncite{{Hall} et~al.}{2007}]{hsher07}
{Hall}, P.~B., {Sadavoy}, S.~I., {Hutsemekers}, D., {Everett}, J.~E., \&
  {Rafiee}, A. 2007, \apj, 665, 174

\bibitem[\protect\astroncite{{Hamann} et~al.}{2008}]{hkrph08}
{Hamann}, F., {Kaplan}, K.~F., {Hidalgo}, P.~R., {Prochaska}, J.~X., \&
  {Herbert-Fort}, S. 2008, \mnras, 391, L39

\bibitem[\protect\astroncite{{Hewett} \& {Foltz}}{2003}]{hf03}
{Hewett}, P.~C. \& {Foltz}, C.~B. 2003, \aj, 125, 1784

\bibitem[\protect\astroncite{{Hewett} et~al.}{1995}]{hfc95}
{Hewett}, P.~C., {Foltz}, C.~B., \& {Chaffee}, F.~H. 1995, \aj, 109, 1498

\bibitem[\protect\astroncite{{Hill} et~al.}{1998}]{hnmtcm98}
{Hill}, G.~J., {Nicklas}, H.~E., {MacQueen}, P.~J., {Tejada}, C., {Cobos
  Duenas}, F.~J., \& {Mitsch}, W. 1998, in Proc. SPIE Vol. 3355, p. 375-386,
  Optical Astronomical Instrumentation, Sandro D'Odorico; Ed., ed.
  S.~{D'Odorico}, vol. 3355 of {\em Presented at the Society of Photo-Optical
  Instrumentation Engineers (SPIE) Conference\/}, 375--386

\bibitem[\protect\astroncite{{Houck} \& {Denicola}}{2000}]{hd00}
{Houck}, J.~C. \& {Denicola}, L.~A. 2000, in ASP Conf. Ser. 216: Astronomical
  Data Analysis Software and Systems IX, eds. N.~{Manset}, C.~{Veillet}, \&
  D.~{Crabtree}, 591--+

\bibitem[\protect\astroncite{{Junkkarinen} et~al.}{2001}]{jsbbchl01}
{Junkkarinen}, V., {Shields}, G.~A., {Beaver}, E.~A., {Burbidge}, E.~M.,
  {Cohen}, R.~D., {Hamann}, F., \& {Lyons}, R.~W. 2001, \apjl, 549, L155

\bibitem[\protect\astroncite{{Kaspi} et~al.}{2000}]{ksnmjg00}
{Kaspi}, S., {Smith}, P.~S., {Netzer}, H., {Maoz}, D., {Jannuzi}, B.~T., \&
  {Giveon}, U. 2000, \apj, 533, 631

\bibitem[\protect\astroncite{{Kinney} et~al.}{1990}]{krk90}
{Kinney}, A.~L., {Rivolo}, A.~R., \& {Koratkar}, A.~P. 1990, \apj, 357, 338

\bibitem[\protect\astroncite{{Knigge} et~al.}{2008}]{ksgc08}
{Knigge}, C., {Scaringi}, S., {Goad}, M.~R., \& {Cottis}, C.~E. 2008, \mnras,
  386, 1426

\bibitem[\protect\astroncite{{Leighly} et~al.}{2009}]{lhcg09}
{Leighly}, K.~M., {Hamann}, F., {Casebeer}, D.~A., \& {Grupe}, D. 2009, \apj,
  701, 176

\bibitem[\protect\astroncite{{Lundgren} et~al.}{2007}]{lwbhsyvb07}
{Lundgren}, B.~F., {Wilhite}, B.~C., {Brunner}, R.~J., {Hall}, P.~B.,
  {Schneider}, D.~P., {York}, D.~G., {Vanden Berk}, D.~E., \& {Brinkmann}, J.
  2007, \apj, 656, 73

\bibitem[\protect\astroncite{{Maronna} et~al.}{2006}]{mmy06}
{Maronna}, R.~A., {Martin}, R.~D., \& {Yohai}, V.~J. 2006, {Robust Statistics:
  Theory and Methods} ({John Wiley \& Sons, Ltd.})

\bibitem[\protect\astroncite{{Murray} et~al.}{1995}]{mcgv95}
{Murray}, N., {Chiang}, J., {Grossman}, S.~A., \& {Voit}, G.~M. 1995, \apj,
  451, 498

\bibitem[\protect\astroncite{{Oke} \& {Gunn}}{1982}]{og82}
{Oke}, J.~B. \& {Gunn}, J.~E. 1982, \pasp, 94, 586

\bibitem[\protect\astroncite{{Pereyra} et~al.}{2006}]{pvthwksb06}
{Pereyra}, N.~A., {Vanden Berk}, D.~E., {Turnshek}, D.~A., {Hillier}, D.~J.,
  {Wilhite}, B.~C., {Kron}, R.~G., {Schneider}, D.~P., \& {Brinkmann}, J. 2006,
  \apj, 642, 87, \eprint{arXiv:astro-ph/0506006}

\bibitem[\protect\astroncite{{Proga} et~al.}{2000}]{psk00}
{Proga}, D., {Stone}, J.~M., \& {Kallman}, T.~R. 2000, \apj, 543, 686

\bibitem[\protect\astroncite{{Ramsey} et~al.}{1998}]{rabbcfggghkklmrrssss98}
{Ramsey}, L.~W., et~al. 1998, in Society of Photo-Optical Instrumentation
  Engineers (SPIE) Conference Series, ed. L.~M. {Stepp}, vol. 3352 of {\em
  Society of Photo-Optical Instrumentation Engineers (SPIE) Conference
  Series\/}, 34--42

\bibitem[\protect\astroncite{{Reichard} et~al.}{2003}]{rrhsvfykb03}
{Reichard}, T.~A., et~al. 2003, \aj, 126, 2594

\bibitem[\protect\astroncite{{Rupke} et~al.}{2002}]{rvs02}
{Rupke}, D.~S., {Veilleux}, S., \& {Sanders}, D.~B. 2002, \apj, 570, 588

\bibitem[\protect\astroncite{{Trump} et~al.}{2006}]{thrrsvkafbkn06}
{Trump}, J.~R., et~al. 2006, \apjs, 165, 1

\bibitem[\protect\astroncite{{Vanden Berk} et~al.}{2004}]{vwkabhirsyblns04}
{Vanden Berk}, D.~E., et~al. 2004, \apj, 601, 692,
  \eprint{arXiv:astro-ph/0310336}

\bibitem[\protect\astroncite{{Vilkoviskij} \& {Irwin}}{2001}]{vi01}
{Vilkoviskij}, E.~Y. \& {Irwin}, M.~J. 2001, \mnras, 321, 4

\bibitem[\protect\astroncite{{Voit} et~al.}{1993}]{vwk93}
{Voit}, G.~M., {Weymann}, R.~J., \& {Korista}, K.~T. 1993, \apj, 413, 95

\bibitem[\protect\astroncite{{Wang} et~al.}{1998}]{wlz98}
{Wang}, T.-G., {Lu}, Y.-J., \& {Zhou}, Y.-Y. 1998, \apj, 493, 1

\bibitem[\protect\astroncite{{Weymann} et~al.}{1991}]{wmfh91}
{Weymann}, R.~J., {Morris}, S.~L., {Foltz}, C.~B., \& {Hewett}, P.~C. 1991,
  \apj, 373, 23

\bibitem[\protect\astroncite{{Wilhite} et~al.}{2005}]{wvkspbrb05}
{Wilhite}, B.~C., {Vanden Berk}, D.~E., {Kron}, R.~G., {Schneider}, D.~P.,
  {Pereyra}, N., {Brunner}, R.~J., {Richards}, G.~T., \& {Brinkmann}, J.~V.
  2005, \apj, 633, 638

\bibitem[\protect\astroncite{{York} et~al.}{2000}]{y+00}
{York}, D.~G., et~al. 2000, \aj, 120, 1579

\end{thebibliography}

\begin{deluxetable}{llllllcc}
\tabletypesize{\scriptsize}
\tablecolumns{8}
\tablewidth{0pc}
\tablecaption{Observation Log\label{obsLogTab}}
\tablehead{\colhead{LBQS Name} & \colhead{$z$}  & \colhead{LBQS Date} & \colhead{Palomar Date} & \colhead{SDSS Date} & \colhead{HET Date} & \colhead{HET Exposure} & \colhead{Max $\Delta t_{sys}$} \\ \colhead{} & \colhead{}  & \colhead{} & \colhead{} & \colhead{} & \colhead{} & \colhead{(min)} & \colhead{(yr)}}
\startdata
$0009+0219$ & 2.642 & 1987 Sep 25 &  &  & 2009 Jan 1 & $20.0$ & $5.8$\\
$0019+0107$ & 2.123 & 1987 Sep 25 & 1988 Aug 9 &  & 2008 Oct 22 & $20.0$ & $6.7$\\
$0021-0213$ & 2.293 & 1988 Sep 9 &  &  & 2007 Dec 5 & $20.0$ & $5.8$\\
$0022+0150$ & 2.770 & 1987 Sep 25 &  &  & 2008 Oct 21 & $17.0$ & $5.6$\\
$0025-0151$ & 2.076 & 1987 Nov 25 &  &  & 2008 Oct 1 & $17.0$ & $6.8$\\
$0029+0017$ & 2.250 & 1987 Nov 25 &  & 2001 Dec 19 & 2008 Oct 21 & $17.0$ & $6.4$\\
$1231+1320$ & 2.380 & 1986 Dec 29 &  &  & 2009 Mar 3 & $20.0$ & $6.6$\\
$1235+0857$ & 2.890 & 1986 Jun 6 & 1988 May 14 &  & 2008 Jun 29 & $17.0$ & $5.7$\\
$1235+1453$ & 2.694 & 1986 Dec 29 & 1988 May 14 & 2005 Mar 13 & 2002 Jun 14 & $15.0$ & $4.9$\\
$1240+1607$ & 2.360 & 1988 May 10 & 1989 Mar 30 &  & 2008 Jun 28 & $20.0$ & $6.0$\\
$1243+0121$ & 2.808 & 1987 Apr 2 & 1989 Mar 31 & 2001 Jan 19 & 2008 Mar 1 & $17.0$ & $5.5$\\
$1314+0116$ & 2.698 & 1988 Mar 16 & 1988 May 14 & 2001 Mar 16 & 2009 Feb 25 & $20.0$ & $5.7$\\
$1442-0011$ & 2.226 & 1988 Mar 16 & 1988 May 13 &  & 2008 Jun 28 & $16.0$ & $6.3$\\
$1443+0141$ & 2.450 & 1988 Mar 16 & 1988 May 14 &  & 2008 Jun 28 & $16.0$ & $5.9$\\

\enddata
\end{deluxetable}

\begin{deluxetable}{lrrr}
\tabletypesize{\scriptsize}
\tablecolumns{4}
\tablewidth{0pc}
\tablecaption{Correlation Tests\label{corrTab}}
\tablehead{\colhead{Comparison} & \colhead{Reference}  & \colhead{Spearman} & \colhead{Number of} \\ \colhead{} & \colhead{Section}  & \colhead{Confidence} & \colhead{Data Pts}}
\startdata
$\Delta EW$ vs. LBQS EW & \S\ref{fracVarEWSec} & $69$ & $14$ \\
$|\Delta$EW$|$ vs. LBQS EW & \S\ref{fracVarEWSec} & $99.62$ & $14$ \\
$\Delta$EW vs. $\langle$EW$\rangle$ & \S\ref{fracVarEWSec} & $ 1$ & $14$ \\
$|\Delta$EW$|$ vs. $\langle$EW$\rangle$ & \S\ref{fracVarEWSec} & $97$ & $14$ \\
$\Delta$EW$_{PL}$ vs. $\Delta$EW$_{HP}$ & \S\ref{varTimeScaleSec} & $68$ & $8$ \\
$\Delta EW/\Delta t$ (PL) vs. $\Delta EW/\Delta t$ (HP) & \S\ref{varTimeScaleSec} & $74$ & $8$ \\
0009+0219: $\Delta S(v)$ vs. $\Delta C(v)$ & \S\ref{siIVCIVShapeVarSec} & $>99.99$ & $166$ \\
1443+0141: $\Delta S(v)$ vs. $\Delta C(v)$ & \S\ref{siIVCIVShapeVarSec} & $>99.99$ & $157$ \\
$\langle $EW$\rangle$ for \ion{Si}{4} vs. \ion{C}{4} & \S\ref{siIVCIVStrengthVarSec} & $62$ & $9$ \\
$\Delta$EW for \ion{Si}{4} vs. \ion{C}{4} & \S\ref{siIVCIVStrengthVarSec} & $23$ & $9$ \\
$|\Delta EW / \langle EW \rangle|$ for \ion{Si}{4} vs. \ion{C}{4} & \S\ref{siIVCIVStrengthVarSec} & $76$ & $9$ \\
$|\Delta EW / \langle EW\rangle|$ for \ion{Si}{4} vs. \ion{C}{4} & \S\ref{siIVCIVStrengthVarSec} & $99.35$ & $8$ \\
~~~(outlier 1235+1453 removed) &  & $$ & $$ \\
LBQS-epoch \ion{C}{4} Emission EW vs. BAL EW & \S\ref{compareEmVarSec} & $56$ & $14$ \\
HET-epoch \ion{C}{4} Emission EW vs. BAL EW & \S\ref{compareEmVarSec} & $41$ & $14$ \\
Change in \ion{C}{4} Emission EW vs. BAL EW & \S\ref{compareEmVarSec} & $66$ & $14$ \\

\enddata
\end{deluxetable}

\begin{deluxetable}{lllll}
\tabletypesize{\scriptsize}
\tablecolumns{5}
\tablewidth{0pc}
\tablecaption{\ion{C}{4} BAL Absorption Equivalent Widths\tablenotemark{a}\label{cIVBALEWTab}}
\tablehead{\colhead{LBQS Name} & \colhead{LBQS}  & \colhead{Palomar} & \colhead{SDSS} & \colhead{HET}}
\startdata
$0009+0219$ & $-12.36 \pm 2.07$ & $$ & $$ & $-30.23 \pm 0.76$\\
$0019+0107$ & $-21.90 \pm 1.23$ & $-26.35 \pm 0.88$ & $$ & $-33.74 \pm 0.84$\\
$0021-0213$ & $-42.91 \pm 1.54$ & $$ & $$ & $-47.78 \pm 0.79$\\
$0022+0150$ & $-15.98 \pm 1.68$ & $$ & $$ & $-10.36 \pm 0.84$\\
$0025-0151$ & $-26.99 \pm 1.41$ & $$ & $$ & $-16.00 \pm 0.91$\\
$0029+0017$ & $-31.99 \pm 1.54$ & $$ & $-33.82 \pm 0.71$ & $-38.02 \pm 0.89$\\
$1231+1320$ & $-27.34 \pm 1.53$ & $$ & $$ & $-33.56 \pm 0.79$\\
$1235+0857$ & $-26.53 \pm 1.42$ & $-28.60 \pm 0.79$ & $$ & $-23.57 \pm 0.88$\\
$1235+1453$ & $-35.64 \pm 3.08$ & $-21.86 \pm 0.86$ & $-39.80 \pm 0.87$ & $-37.94 \pm 0.97$\\
$1240+1607$ & $-28.14 \pm 1.48$ & $-36.30 \pm 0.94$ & $$ & $-34.40 \pm 0.79$\\
$1243+0121$ & $-41.77 \pm 2.53$ & $-40.38 \pm 0.79$ & $-33.63 \pm 0.70$ & $-36.85 \pm 0.78$\\
$1314+0116$ & $-22.14 \pm 2.37$ & $-22.08 \pm 0.90$ & $-30.22 \pm 0.73$ & $-42.10 \pm 0.73$\\
$1442-0011$ & $-37.21 \pm 1.24$ & $-33.48 \pm 0.83$ & $$ & $-39.76 \pm 0.77$\\
$1443+0141$ & $-43.99 \pm 1.56$ & $-48.83 \pm 0.81$ & $$ & $-44.86 \pm 0.76$\\

\enddata
\tablenotetext{a}{Equivalent widths are given in Angstroms.}
\end{deluxetable}

\begin{deluxetable}{lllll}
\tabletypesize{\scriptsize}
\tablecolumns{5}
\tablewidth{0pc}
\tablecaption{\ion{Si}{4} And \ion{C}{4} BAL Absorption Equivalent Widths (Out To $-20,000$~km~s$^{-1}$)\tablenotemark{a}\label{siCIVEW20Tab}}
\tablehead{\colhead{LBQS Name} & \colhead{LBQS \ion{Si}{4}}  & \colhead{LBQS \ion{C}{4}} & \colhead{HET \ion{Si}{4}} & \colhead{HET \ion{C}{4}}}
\startdata
0009+0219 & $-0.79 \pm 1.28$ & $-7.74 \pm 1.83$ & $-6.45 \pm 0.69$ & $-23.67 \pm 0.61$\\
0022+0150 & $-3.58 \pm 1.14$ & $-11.01 \pm 1.44$ & $-3.05 \pm 0.68$ & $-6.09 \pm 0.70$\\
1231+1320 & $-29.32 \pm 1.04$ & $-10.31 \pm 1.32$ & $-21.05 \pm 0.66$ & $-14.64 \pm 0.70$\\
1235+0857 & $-5.15 \pm 1.00$ & $-22.39 \pm 1.15$ & $-5.55 \pm 0.74$ & $-21.20 \pm 0.68$\\
1235+1453 & $-1.08 \pm 1.95$ & $-29.63 \pm 2.73$ & $-8.23 \pm 0.87$ & $-29.55 \pm 0.79$\\
1240+1607 & $-11.53 \pm 1.08$ & $-25.40 \pm 1.20$ & $-9.96 \pm 0.73$ & $-30.24 \pm 0.59$\\
1243+0121 & $-8.90 \pm 1.59$ & $-39.92 \pm 2.11$ & $-12.12 \pm 0.66$ & $-32.24 \pm 0.60$\\
1314+0116 & $-3.81 \pm 1.36$ & $-21.97 \pm 2.06$ & $-9.55 \pm 0.68$ & $-34.93 \pm 0.56$\\
1443+0141 & $-12.00 \pm 1.19$ & $-32.29 \pm 1.27$ & $-12.46 \pm 0.74$ & $-33.00 \pm 0.61$\\

\enddata
\tablenotetext{a}{LBQS and HET equivalent widths measured from 0~to $-20,000$~km~s$^{-1}$, given in Angstroms.}
\end{deluxetable}

\begin{deluxetable}{lllll}
\tabletypesize{\scriptsize}
\tablecolumns{5}
\tablewidth{0pc}
\tablecaption{\ion{C}{4} Emission Equivalent Widths\tablenotemark{a}\label{cIVEmEWTab}}
\tablehead{\colhead{LBQS Name} & \colhead{LBQS}  & \colhead{Palomar} & \colhead{SDSS} & \colhead{HET}}
\startdata
$0009+0219$ & $18.23$ & $$ & $$ & $9.79$\\
$0019+0107$ & $32.51$ & $26.79$ & $$ & $29.87$\\
$0021-0213$ & $17.48$ & $$ & $$ & $16.67$\\
$0022+0150$ & $31.58$ & $$ & $$ & $23.49$\\
$0025-0151$ & $34.11$ & $$ & $$ & $22.36$\\
$0029+0017$ & $58.59$ & $$ & $44.79$ & $50.35$\\
$1231+1320$ & $22.76$ & $$ & $$ & $8.92$\\
$1235+0857$ & $55.72$ & $25.92$ & $$ & $29.89$\\
$1235+1453$ & $22.35$ & $33.56$ & $45.01$ & $35.25$\\
$1240+1607$ & $29.53$ & $37.00$ & $$ & $19.03$\\
$1243+0121$ & $44.80$ & $39.46$ & $45.44$ & $46.15$\\
$1314+0116$ & $57.99$ & $53.78$ & $41.11$ & $31.35$\\
$1442-0011$ & $20.88$ & $28.27$ & $$ & $25.00$\\
$1443+0141$ & $22.77$ & $29.92$ & $$ & $21.19$\\

\enddata
\tablenotetext{a}{All EWs are given in Angstroms.  We do not include formal errors because the uncertainty is primarily a factor of our manual adjustments to generate consistent fits for a variety of complex and absorbed line profiles (\S\ref{compareEmVarSec}).}
\end{deluxetable}


\clearpage

\clearpage

\clearpage

\clearpage

\clearpage

\clearpage

\clearpage

\clearpage

\clearpage

\clearpage

\clearpage

\clearpage

\end{document}